\documentclass[a4paper,11pt]{article}
\pdfoutput=1
\usepackage{graphicx,amssymb,amstext,amsmath}

\usepackage{color}
\usepackage{hyperref}

\newtheorem{rema}{Remark}[section]

\newtheorem{propo}{Proposition}[section]

\newtheorem{conjecture}{Conjecture}[section]

\newcommand{\bc}{\begin{center}}
\newcommand{\ec}{\end{center}}
\def\ba#1{\begin{array}{#1}\displaystyle}
\newcommand{\ea}{\end{array}}
\newcommand{\z}{\\[4mm] \displaystyle}

\newcommand{\beq}{\begin{equation}}
\newcommand{\eeq}{\end{equation}}
\newcommand{\beqa}{\begin{eqnarray}}
\newcommand{\eeqa}{\end{eqnarray}}
\newcommand{\no}{\nonumber}
\newcommand{\n}{\nonumber\\}
\newcommand{\bi}{\begin{itemize}}
\newcommand{\ei}{\end{itemize}}
\def\mato#1{\left(\ba{#1}} 
\def\matf{\ea\right)}

\def\lt#1{\left#1}
\def\rt#1{\right#1}
\def\t#1{\tilde{#1}}

\def\b#1{\bar{#1}}
\def\frc#1#2{\frac{#1}{#2}}

\newcommand{\p}{\partial}

\newcommand{\vac}{{\rm vac}}
\newcommand{\bra}{\langle}
\newcommand{\ket}{\rangle}
\newcommand{\Z}{{\mathbb{Z}}}
\newcommand{\N}{{\mathbb{N}}}
\newcommand{\R}{{\mathbb{R}}}
\newcommand{\C}{{\mathbb{C}}}

\newcommand{\Or}{{\cal O}}

\newcommand{\ep}{\epsilon}
\newcommand{\varep}{\varepsilon}

\newcommand{\Tr}{{\rm Tr}}

\newcommand{\Res}{{\rm Res}}
\newcommand{\End}{{\rm End}}

\newcommand{\liou}{{\cal L}}
\newcommand{\la}{{\bf a}}
\newcommand{\lb}{{\bf b}}
\newcommand{\lH}{{\bf H}}
\newcommand{\lP}{{\bf P}}
\newcommand{\lU}{{\bf U}}

\newcommand{\nl}{\mbox{\tiny ${\circ\atop\circ}$}}

\date{April 2011}
\title{Correlation functions of twist fields from Ward identities in the massive Dirac theory}
\author{Benjamin Doyon \and James Silk}

\begin{document}

\begin{titlepage}
\vspace{0.2cm}
\begin{center}

{\Large {\bf Form factors in equilibrium and non-equilibrium mixed states of the Ising model}}

\vspace{0.8cm} {\large \text{Yixiong Chen and Benjamin Doyon}}

\vspace{0.2cm}
{Department of Mathematics, King's College London, Strand WC2R 2LS, UK }
\end{center}

\vspace{1cm} 
Using the ``Liouville space'' (the space of operators) of the massive Ising model of quantum field theory, there is a natural definition of form factors in any mixed state. These generalize the usual form factors, and are building blocks for mixed-state correlation functions. We study the cases of mixed states that are diagonal in the asymptotic particle basis, and obtain exact expressions for all mixed-state form factors of order and disorder fields. We use novel techniques based on deriving and solving a system of nonlinear functional differential equations. We then write down the full form factor expansion for mixed-state correlation functions of these fields. Under weak analytic conditions on the eigenvalues of the density matrix, this is an exact large-distance expansion. The form factors agree with the known finite-temperature form factors when the mixed state is specialized to a thermal Gibbs ensemble. Our results can be used to analyze correlation functions in generalized Gibbs ensembles (which occur after quantum quenches). Applying this to the density matrix for non-equilibrium steady states with energy flows, we observe that non-equilibrium form factors have branch cuts in rapidity space. We verify that this is in agreement with a non-equilibrium generalization of the KMS relations, and we conjecture  that the leading large-distance behavior of order and disorder non-equilibrium correlation functions contains oscillations in the log of the distance between the fields.
\vfill

{\ }\hfill March 2014

\end{titlepage}

\tableofcontents

\section{Introduction}

The exact evaluation of correlation functions of local fields is one of the main challenges of quantum field theory (QFT). Correlation functions are of crucial importance: they encode, in principle, all physical information, and they are immediately related to experimental results in condensed matter systems near criticality.

A particularly interesting family of models of QFT are 1+1-dimensional massive integrable models \cite{Karow,Zamo,Smirnov,Delfino} (see the book \cite{Mussardo}). Their extended symmetries allow for exact evaluations of many quantities and give rise to a rich mathematical structure to build upon. They have also found a multitude of applications to modern condensed matter physics \cite{Ess}. Much progress has been achieved in the past 40 years in evaluating vacuum correlation functions in integrable QFT, in particular via the factorized scattering theory and the associated theory of form factor expansions \cite{Karow,Zamo,Smirnov} (K\"allen-Lehmann expansions). The large-distance (or low-energy) structure of two-point functions is well understood via the analytic structure of form factors, and both their large-distance and short-distance behaviors are under good control (see for instance \cite{ZamoLee,Yurov}), giving agreement with expected general principles of QFT.

In recent years, due to developments in various areas of theoretical physics and to new experimental results, interest has grown for an accurate evaluation of correlation functions in general mixed states. Besides thermal Gibbs states, these also include generalized Gibbs ensembles (GGEs) believed to occur after quantum quenches in integrable models \cite{Rigol,Rigol2} (see the review \cite{Pol} and the analytical results confirming this in the Ising model \cite{Fagotti, Fagotti1}), non-equilibrium energy-carrying quantum steady states \cite{Aschbacher2,Ben4,BDaihp,Ben2,DeLuca}), and others. In most of the physical applications of current interest, the density matrix is diagonal in the basis of asymptotic states and multiplicative on the particles (we will refer to these as diagonal mixed states). There is still relatively little known about the structure of two-point functions in such general situations, and how this structure depends on the mixed state. Correlation functions in thermal states (or on the cylinder) have been widely studied in general QFT (see for instance the books \cite{Kapusta}), and with more precision in massive integrable QFT \cite{Ess,Leclair,Sachdev1,Sachdev2,Leclair2,Saleur,Mussardo2,Konik,Fonseca1,Fonseca2,Ben1,Altshuler,Reyes06,Damle05,Rapp,Ben3,Gamsa,Pozsgay1,Pozsgay2,EssKon1,EssKon2,Pozsgay3,Sz}. The structure is relatively well understood, although still much work needs to be done in integrable QFT in order to get as powerful large-distance or large-time expansions as for vacuum two-point functions. The study of vacuum expectation values and correlation functions in generalized Gibbs ensembles is however much more recent. Analytic leading-asymptotic results for two-point functions are available in the Ising model \cite{Fagotti,Fagotti2} and series expansions for vacuum expectation values in GGEs of general integrable models \cite{MusQuench}. For correlation functions in non-equilibrium energy-carrying steady states there are even fewer results, notably a leading-decay result at large distances in the XY spin chain \cite{Aschbacher}. Hence, exact QFT expansions for general diagonal mixed states would shed much light onto the structure of correlation functions.

In the present paper, we obtain for the first time exact, convergent large-distance (space-like) expansions for real-time correlation functions of order and disorder fields in the Ising model of QFT in general diagonal mixed states.

The Ising model is a paradigmatic model of integrable QFT, and provides a tool to benchmark many new exact method or new physical context. It describes the scaling limit of the anisotropic XY quantum spin chain (including the Ising quantum chain) near the critical value $h_c$ of the external transverse magnetic field $h$. Order and disorder fields in the QFT model describe the scaling limit of the order parameter in the ordered ($h>h_c$) and disordered ($h<h_c$) regimes, respectively, asymptotically closed to the critical value. Although the QFT model has trivial asymptotic particle theory -- it is a free massive relativistic Majorana fermion theory -- the order and disorder fields are non-local with respect to the asymptotic particles, whence are ``interacting'' fields. Therefore, their two-point function should already contain  part of the non-trivial structure of two-point functions in (integrable) models of interacting particles.

The large-distance expansion we obtain is convergent and gives in principle a numerically efficient expression for correlation functions in a large family of mixed states, including generalized Gibbs ensembles. It can be used to analyze their large-distance asymptotic expansion: for instance, we find agreement with leading large-distance results of \cite{Fagotti2}, and we compute new subleading terms. In non-equilibrium steady states our expansion requires further regularization which we do not study in detail here, but the leading large-distance behavior is in agreement with the results of \cite{Aschbacher}, and we conjecture the presence of logarithmic oscillating subleading factors.

We use the method of form factor expansion with respect to the mixed-state ``vacuum'' in the Liouville space (essentially the space of operators on the Hilbert space). The Liouville space construct is an old idea \cite{Fano1,Crawford,Fano2,Schmutz,Umezawa,Arimitsu} which found applications in thermal and non-equilibrium physics, and which is ultimately based on the even older GNS construction \cite{Gelfand,Segal} of $C^*$-algebras (see for instance the book \cite{Arveson}). To our knowledge, it was for the first time applied to massive integrable models of QFT in \cite{Ben1}. In the present paper, generalizing the thermal situation of \cite{Ben1,Ben3}, we define the notion of form factors in general diagonal mixed states, we calculate them explicitly in the Ising model, and we explain how they can be used to obtain large-distance expansions of two-point functions.

\subsection{Main result}

Our main result is as follows. Let $m$ be the mass of the asymptotic particle of the Ising model, and $\prod_{j=1}^N e^{- W(\theta_j)}$ be the eigenvalue of the density matrix on the $N$-particle state with rapidities $\{\theta_j\}$.  Let $G(x,t)$ (resp. $\t G(x,t)$) be the scaling limit of the two-point functions of the order parameter in the ordered (resp. disordered) regime of the Ising model. If $W(\theta)$ is analytic on a neighborhood of the real line, then
\begin{eqnarray}
\lefteqn{G(x,t) \pm \t G(x,t)}\n
&\propto&
e^{-|x|{\cal E}}\,\sum_{N=0}^{\infty} \frc{(\pm 1)^N}{N!}
\sum_{\ep_1,\ldots,\ep_N \in \{+,-\}}\int
\lt(\prod_{j=1}^N d\theta_j \frc{i^{\ep_j}\,
\exp\lt[K_{\ep_j}(\theta_j)+i\ep_jp_{\theta_j} |x| - i\ep_jE_{\theta_j} t\rt]}{
	2\pi i\lt(1-e^{-\ep_j W(\theta)}\rt)}\rt)
\times \n && \times\; \prod_{j<k} \lt(\tanh
\frc{\theta_j-\theta_k +i(\ep_j-\ep_k)}2\rt)^{2\ep_j\ep_k}\label{ft}
\label{corrsimple}
\end{eqnarray}
where
\beq\label{pE}
	p_\theta = m\sinh\theta,\quad E_\theta = m\cosh\theta,
\eeq
and
\beqa\label{E}
	{\cal E}
	&=& \int_{-\infty}^\infty \frc{d\theta}{2\pi}
	m\cosh\theta \log\lt(
	\coth\frc{W(\theta)}2
	\rt)\\
\label{K}
	K_\ep(\theta) &=& 
	\ep
	\int_{-\infty}^{\infty}
	\frac{d\theta'}{\pi i}
	\frac1{\sinh(\theta-\theta'+i\ep{\bf 0})}
	\log\left(
	\coth \frc{W(\theta')}2
	\right)
\eeqa
(here and below, ${\bf 0}$ is a positive infinitesimal). The proportionality factor in (\ref{ft}) is $x$ and $t$ independent, but depends on $W$. The rapidity contours in (\ref{ft}) can be shifted in the direction $i\ep_j$, up to the position of singularities which depend on the analytic properties of $W(\theta)$. These singularities determine the large-distance asymptotic expansion. For instance, the disordered-regime two-point function decays proportionally, up to algebraic or other non-exponential factors, to $e^{-|x|({\cal E}+ m|{\rm Im}(\sinh \theta^\star)|)}$ as $x\to\infty$, where $\theta^\star$ is the minimum of $|{\rm Im}(\sinh\theta)|$ over the singularities in $\theta$ of $\lt(1-e^{\pm W(\theta)}\rt)^{-1}$. The presence and type of extra factors is determined by the type of singularity (pole, branch point, etc.). Similar statements hold for higher number of particles, giving, in principle, the complete asymptotic expansion.

\subsection{Calculation methods}

In order to evaluate the usual form factors in the vacuum, the most commonly used method in integrable QFT is that by which one solves a Riemann-Hilbert problem in the (analytically continued) rapidities of the asymptotic particles \cite{Smirnov}. A similar Riemann-Hilbert problem was derived in \cite{Ben1,Ben3} for thermal form factors in the Ising model, in particular putting in a Liouville-space set-up a derivation for form factors on the circle in \cite{Fonseca2}. In the present case of general mixed states, however, these techniques cannot be applied, because we cannot assume any strong analytic properties of the density matrix.

Another way, as explained in \cite{Ben1} (see also \cite{Altshuler}), is to realize that thermal form factors are ``analytic continuation'' of vacuum form factors in the quantization on the circle, and thermal form factor expansions give the real-time version of form factor expansions on the circle (compactified imaginary time). Form factors on the circle in the Ising model were first evaluated from lattice form factors in free-fermion models on the cylinder \cite{Bugrij,Bugrij2,Bugrij3,Bugrij4,Bugrij5,Iorgov,Iorgov2}, and can be evaluated directly in the Ising field theory \cite{Fonseca1,Fonseca2}. Again in the present case of general mixed states, these techniques cannot be applied because there is (yet) no clear geometric picture behind general density matrices that would lead to alternative quantization schemes (but see Subsection \ref{ssectfurther} for ideas concerning a relation with the analytic structure of the eigenvalues of the density matrix).

Instead, our calculation uses novel methods, to our knowledge not used before in the context of integrable QFT: we derive and solve a system of bilinear first-order functional differential equations for mixed-state form factors. We observe that our mixed-state form factors have in general a reduced analyticity region in rapidity space as compared to vacuum form factors of integrable models. In particular, in the case of the non-equilibrium energy-carrying steady states, the form factors have branch points at zero rapidity.

It is worth mentioning that, like in \cite{Ben1,Ben3}, our mixed-state form factor method {\em does not} require any re-summation of partition-function divergencies, involving or not finite-volume regularization. These re-summations are required when using the standard {\em vacuum} form factors in order to evaluate, for instance, thermal averages in integrable QFT by explicitly performing traces (there is a quite developed technology on this subject \cite{Leclair,Leclair2,Saleur,Mussardo2,Konik,Pozsgay1,Pozsgay2,EssKon1,EssKon2,Pozsgay3,Sz}), or when studying quantum quenches \cite{Schuricht2012}. In our framework, re-summations are automatically and unambiguously performed in the definition of mixed-state form factors (up to their overall normalization), and in the resulting form-factor expansion. It is also worth mentioning that our method is a {\em real-time} method, something which is necessary since as we mentioned, by opposition to thermal Gibbs states, it is not expected that there be any clear geometric principle in imaginary time for general diagonal mixed state. Our mixed-state form factor method is, in principle, generalizable to interacting integrable QFT (although explicit form-factor calculations need more developments).

\subsection{Organization of the paper}

The paper is organized as follows. In Section 2 we provide short overview of basic facts concerning the Ising model (scaling limit and field theory). In section 3, we introduce basic notions concerning the Liouville space and we define mixed-state form factors. In Section 4 we present our main results, including the mixing phenomenon for normal-ordered product of fermion fields, the exact mixed-state form factors of order and disorder fields, and the form factor expansion for mixed-state correlation functions. In Section 5 we discuss some possible applications of our results: quantum quenches and non-equilibrium thermal-flow steady states. In Section 6 we present calculations showing and explaining the main results. Finally we conclude in Section 7. Appendices give technical details on various aspects.

\section{Ising model}

\subsection{Ising and XY quantum chains in the scaling limit}

Consider the infinite-length quantum chain described by the hamiltonian (the generic $XY$ model \cite{Lieb61,Katsura62})
\beq\label{Hchain}
	H_{\rm chain} = -\frc J2 \sum_n \lt[
	\frc{1+\kappa}2 \sigma_n^x \sigma_{n+1}^x +
	\frc{1-\kappa}2 \sigma_n^y\sigma_{n+1}^y
	+h\sigma_n^z\rt]
\eeq
where $\sigma_n^i$ are Pauli matrices at site $n$, $\kappa\in[-1,1]$ is the anisotropy parameter (in the following we will exclude the value $\kappa=0$), $h$ is the (dimensionless) external transverse magnetic field and $J>0$ is an energy coupling constant. The automorphism $(\sigma^{x},\sigma^y,\sigma^z)\mapsto (\sigma^y,\sigma^x, -\sigma^z)$ corresponds to $(\kappa,h)\mapsto(-\kappa,-h)$, hence it is sufficient to consider the region $h\geq0$. At the special values $\kappa=\pm1$ this is the quantum Ising model in a transverse field. This model is in general integrable: it is equivalent to a free-fermion model via the standard Jordan-Wigner transformation \cite{Lieb61,Katsura62,JordanWigner}. Its phase diagram and critical behavior have been described exactly \cite{McCoy68,Pfeuty70,BarouchMcCoyII}.

The spectrum is described by fermionic excitation modes of energies
\[
	\ep_q = J\sqrt{(1-\kappa^2)\cos^2 q - 2h\cos q + \kappa^2+h^2}
\]
for $q\in[-\pi,\pi)$. We see that the model is gapless at the critical value $h= 1$. This critical point is in the Ising universality class for every $\kappa\neq 0$, hence it is described by the Ising model of conformal field theory with central charge $c=1/2$ \cite{Belavin}: the free field theory of massless Majorana fermions\footnote{Often, one only takes a particular locality sector of this theory, which excludes the Majorana fermions themselves.}. When $h$ approaches the critical value $1$, the correlation length $\xi$, measured in number of sites, diverges, $\xi \sim  |\kappa|/|h-1|$. The low-energy excitations are then slowly varying over the chain and are correctly described, in the limit, by arbitrary numbers of modes with a relativistic dispersion relation (that is, it is consistent to take an arbitrary number of modes and yet take the low-$q$ form of $\ep_q$). One can construct smooth relativistic fields from which the low-energy and large-distance universal behaviors can be predicted. This gives the massive Majorana relativistic field theory (see for instance \cite{Itzykson}), with Hamiltonian
\beq\label{Hma}
	H = \int dx\,:\lt[-\frc{v_F}2\,i\psi(x)\p_x\psi(x) +\frc{v_F}2\,i\b\psi(x)\p_x\b\psi(x) + m\,i\psi(x)\b\psi(x)\rt]:
\eeq
formed out of mutually anti-commutating fields $\psi(x)$ and $\b\psi(x)$ satisfying the anti-commutation relations
\beq\label{carpsi}
	\{\psi(x),\psi(x')\}=\{\b\psi(x),\b\psi(x')\}=\delta(x-x').
\eeq
As usual, in (\ref{Hma}) the normal ordering corresponds to the energy of the vacuum state $|\vac\ket$ being set to zero. Here $v_F = |\kappa| Ja$ is the velocity of the excitations and $m = J|h-1|/v_F^2$ is their mass; the constant $a$ is the lattice spacing, introduced to provide distance units, an ``infinitesimal'' scale over which $\psi(x)$ is essentially constant.

This field theory will describe correctly physical quantities at energy scales $T\lesssim mv_F^2= J|h-1|$ and dimensionful length scales $\ell \gtrsim a\xi = a|\kappa|/|h-1|$, in the limit where $h\to 1$. For instance, the energy scale $T$ may be the temperature, and the length scale $\ell$ the distance between local fields in a correlation function. The full relation is obtained by identifying appropriate fields in the Majorana theory with observables in the quantum chain. This identification depends on the regime considered: either $h<1$ or $h>1$, and either $\kappa>0$ or $\kappa<0$. For simplicity, in the following we will restrict ourselves to the region
\[
	\kappa\in(0,1]
\]
and to the observables $\sigma^x_n$ and $\sigma^z_n$.

These observables $\sigma^x_n$ and $\sigma^z_n$ give rise, in the scaling limit, to various local fields in the Majorana theory. Three fields are of importance: two primary twist fields $\sigma$ and $\mu$ associated to the $\Z_2$ symmetry
\beq\label{Z2}
	(\psi,\b\psi)\mapsto (-\psi,-\b\psi),
\eeq
and the field
\beq
	\varep =\, i:\b\psi\psi:.
\eeq
The twist field $\sigma$, the ``order'' field, only generates, from the vacuum, even numbers of particles, while the twist field $\mu$, the ``disorder'' field, only generates odd numbers of particles; the former has a nonzero expectation value, contrary to the latter. These are fields representing the end-point of cuts through which other fields are affected by the symmetry transformation (\ref{Z2}). There are two directions the cut can go on the chain: towards the right, or towards the left. With the information of the direction of the cut, we will denote the twist fields by $\sigma_+$, $\mu_+$, and $\sigma_-$, $\mu_-$ (for cuts towards the right and the left, respectively).

The identification is as follows:
\beq\label{idsl}
	\ba{c|c|c} & h\to1^+ & h\to1^-  \\ \hline
	\sigma^z_n - \bra\vac|\sigma^z_n|\vac\ket & 2\varep(x) &-2\varep(x)
	\\
	\sigma^x_n & \mu_+(x) & \sigma_+(x)
	\ea
\eeq
Here we have chosen a direction for the cut; this choice is arbitrary, but must be kept the same for all fields in a correlation function. As it should, in the low-$h$ regime $h\to 1^-$, $\sigma^x_n$ takes a nonzero expectation value (there is ``order''), while in the large-$h$ regime $h\to 1^+$, its expectation value is zero. The identification implies that, for instance, in a state $\rho$ of characteristic energy scale $T$,
\beq\label{sl}
	\lim_{h\to1} (m\xi)^{2d}\bra \sigma_n^x \sigma_{n'}^x\ket_\rho^{\rm chain} \propto
	\lt\{\ba{ll} \bra\sigma_+(x)\sigma_+(y)\ket_\rho & \mbox{(ordered regime)} \\
	 \bra\mu_+(x)\mu_+(y)\ket_\rho & \mbox{(disordered regime)}
	 \ea \rt.
\eeq
where $d=1/8$ is a universal quantity, the scaling dimension of $\sigma_\pm$ and $\mu_\pm$. On the right-hand side, correlation functions are evaluated in the QFT. The limit is taken with $x=an$, $y=an'$, $n-n'=\delta\xi$ and $T=\gamma m v_F^2$ for some fixed constants $\delta,\gamma$. Here the QFT fields are both evaluated at time 0, but a similar relation is often expected to hold for time-evolved operators. Besides an overall dimensionful factor $m^{2d}$, the QFT correlation function only depends on dimensionless combinations that do not involve the microscopic energy scale $J$.  In a translation-invariant state, these are the finite numbers $(x-y)T/v_F=\delta\gamma$ and $(x-y)mv_F = \delta$. That is, we have for instance $\bra\sigma_+(x)\sigma_+(y)\ket_\rho^{\rm QFT} = m^{2d} f(\gamma,\delta)$, where the function $f(\gamma,\delta)$ is a {\em universal scaling function}. The proportionality constant in (\ref{sl}) is non-universal and depends in general on $\kappa$.

With time-evolved field $\Or(x,t) = e^{iHt}\Or(x) e^{-iHt}$, we will denote the twist-field correlation functions by
\beq\label{GGt}
	G(x,t) = \bra\sigma_+(x,t)\sigma_+(0,0)\ket_\rho ,\quad
	\t G(x,t) = \bra\mu_+(x,t)\mu_+(0,0)\ket_\rho.
\eeq
In the following, we set $v_F=1$.

\subsection{Form factors and correlation functions}

The evaluation of correlation functions of interacting fields in massive relativistic QFT is a rather complicated task. In the context of integrable QFT, one may use factorized scattering theory, based on the fact that the scattering matrix factorizes into two-particle scattering matrices \cite{Karow,Zamo,Smirnov,Delfino} (see also the book \cite{Mussardo}). Then, a very powerful method for vacuum correlation functions is that using the spectral decomposition and the exact evaluation of matrix elements involved \cite{Karow,Berg,Smirnov}. For a QFT with a spectrum of $K$ particle types, the set of asymptotic states (say $in$ states) is described by giving the number of particles $N\in\N$ in the state, and the associated rapidities $\theta_j\in\R$ and particle types $\ep_j\in\{1,\ldots,K\}$,
\[
	|\theta_1,\ldots,\theta_N\ket_{\ep_1,\ldots,\ep_N},\quad
	\theta_1>\cdots>\theta_N.
\] 
The spectral decomposition follows from the statement that the basis of asymptotic states can be used to resolve the identity. From standard relativistic and translation invariance, correlation functions in the vacuum at space-like distances $r=\sqrt{x^2-t^2}>0$ are then expressed as infinite series:
\beq\label{ff0exp}
	\bra\vac|\Or^\dag(x,t)\Or(0,0)|\vac\ket
	= \sum_{N=0}^\infty\sum_{\ep_1,\ldots,\ep_N}
	\int_{\theta_1>\cdots>\theta_N}\hspace{-1cm}
	d\theta_1\cdots d\theta_N
	|\bra\vac|\Or|\theta_1,\ldots,\theta_N\ket_{\ep_1,\ldots,\ep_N}|^2
	e^{-\sum_{j} rm_{\ep_j}\cosh\theta_j}.
\eeq

In (\ref{ff0exp}), the term with $k=0$ is simply $|\bra\vac|\Or|\vac\ket|^2$, and in $\bra\vac|\Or|\theta_1,\ldots,\theta_N\ket_{\ep_1,\ldots,\ep_N}$ the local field $\Or$ is implicitly at the space-time point $(0,0)$. This expansion is believed to form a convergent series. Form factors $f^\Or_{\ep_1,\ldots,\ep_N}(\theta_1,\ldots,\theta_N)$ in the context of massive integrable models of QFT are analytic extensions of matrix elements of local fields $\bra\vac|\Or|\theta_1,\ldots,\theta_N\ket_{\ep_1,\ldots,\ep_N}$ and can be evaluated exactly by solving a system of equations and analytic properties, the so-called form-factor equations \cite{Karow,Berg,Smirnov}. In particular, the analytic continuation from the region $\theta_1>\ldots>\theta_N$, where they correspond to $in$ states, to regions with different orderings, gives rise the two-particle scattering matrix (Watson's lemma).

The spectrum of the Ising model contains only one particle type. The space of $in$ states is a Fock space ${\cal H}$ over the canonical anti-commutation relations
\beq\label{can}
	\{a(\theta),a(\theta')\}=0,\quad
	\{a(\theta),a^\dag(\theta')\} = \delta(\theta-\theta')
\eeq
with the asymptotic states identified as
\[
	|\theta_1,\ldots,\theta_N\ket = a^\dag(\theta_1)\cdots a^\dag(\theta_N)|\vac\ket
\]
(we will denote $|\theta_1,\ldots,\theta_N\ket^\dag = \bra\theta_1,\ldots,\theta_N|$). The canonical algebra is in agreement with the analytic continuation of form factors to different orderings: the two-particle scattering matrix is $-1$. Naturally, the free fermion fields of the Ising model $\psi(x,t)$ and $\b\psi(x,t)$ can be expressed linearly in terms of the operators $a(\theta)$ and $a^\dag(\theta)$:
\beqa
	\psi(x,t) &=& \frc12 \sqrt{\frc m\pi} \int d\theta\,e^{\theta/2}
	\lt(a(\theta)\,e^{ip_\theta x - i E_\theta t}
	+ a^\dag(\theta)\,e^{-ip_\theta x + i E_\theta t}\rt) \n
	\b\psi(x,t) &=& -\frc i2 \sqrt{\frc m\pi} \int d\theta\,e^{-\theta/2}
	\lt(a(\theta)\,e^{ip_\theta x - i E_\theta t}
	- a^\dag(\theta)\,e^{-ip_\theta x + i E_\theta t}\rt)
	\label{psimodes}
\eeqa
where $p_\theta$ and $E_\theta$ (see \eqref{pE}) are the relativistic momentum and energy associated to the rapidity $\theta$ and with mass $m$. The Hamiltonian and momentum operators are
\beq
	H = \int d\theta\, E_\theta\,a^\dag(\theta) a(\theta),\quad
	P = \int d\theta\,p_\theta\,a^\dag(\theta) a(\theta).
\eeq

The fields $\sigma_\pm$ and $\mu_\pm$ in the Ising model are twist fields associated to the $\Z_2$ symmetry (\ref{Z2}). Since $\sigma_\pm$ couple states differing by even numbers of particles only, it is of bosonic statistics; while $\mu_\pm$ are of fermionic statistics. Their twist conditions, including the statistics, are a direct consequence of the Jordan-Wigner transformation, and are given by
\beq\label{twist}
	\sigma_\pm(x)\psi(x') = \lt\{\ba{ll} \pm\psi(x')\sigma_\pm(x) & (x>x') \\
	\mp\psi(x')\sigma_\pm(x) & (x<x')\ea\rt.,\quad
	\mu_\pm(x)\psi(x') = \lt\{\ba{ll} \mp\psi(x')\mu_\pm(x) & (x>x') \\
	\pm\psi(x')\mu_\pm(x) & (x<x')\ea\rt.
\eeq
and similar conditions hold with the replacement $\psi\mapsto\b\psi$. 

Along with the conditions that $\sigma_+$ and $\mu_+$ be of the lowest scaling dimension, it is expected that \eqref{twist} uniquely define these fields up to normalization. Once they have been defined and normalized, one can construct $\sigma_-$ and $\mu_-$ unambiguously using the unitary operator
\beq\label{Z}
	Z:=\exp\lt[i\pi \int d\theta \,a^\dag(\theta)a(\theta)\rt].
\eeq
This operator implements the $\Z_2$ symmetry transformation \eqref{Z2}, and we define
\beq\label{sZ}
	 \sigma_-(x,t) = \sigma_+(x,t)Z,\quad \mu_-(x,t) = \mu_+(x,t)Z.
\eeq
This is in agreement with (but not uniquely fixed by) \eqref{twist}, and the multiplication by $Z$ does not modify the scaling dimension.

Although the Ising model is a free-fermion model, hence has a trivial scattering matrix, the order and disorder fields $\sigma_\pm$ and $\mu_\pm$ are not ``free field''. That is, they have nontrivial form factors for all (even or odd, respectively) particle numbers (contrary to the field $\varep$). The nonzero form factors of the fields $\sigma_\pm$, $\mu_\pm$  \cite{Berg,Yurov,Ben1} and $\varep$ are given by
\beqa
	f^{\sigma_\pm}(\theta_1,\ldots,\theta_N) &\stackrel{N\ {\rm even}}=&
	\lt(\frc{i}{\sqrt{2\pi}}\rt)^N \;\bra\sigma\ket
	\prod_{1\leq i<j\leq N} \tanh\lt(\frc{\theta_j-\theta_i}2\rt) \n
	f^{\mu_\pm}(\theta_1,\ldots,\theta_N) &\stackrel{N\ {\rm odd}}=&
	\pm\frc1{\sqrt{i}}\lt(\frc{i}{\sqrt{2\pi}}\rt)^N \;\bra\sigma\ket
	\prod_{1\leq i<j\leq N} \tanh\lt(\frc{\theta_j-\theta_i}2\rt) \n
	f^\varep(\theta_1,\theta_2) &=& -\frc m{2\pi}
	\sinh\lt(\frc{\theta_2 -\theta_1}2\rt)\label{ffvac}
\eeqa
where $\bra\sigma\ket :=\bra\vac|\sigma_\pm|\vac\ket$ is the vacuum expectation value. Note that all other matrix elements can be evaluated by using crossing relations \cite{Karow,Berg,Smirnov,Yurov}, and that \cite{Ben1}
\beq\label{herm}
 	\sigma_\pm^\dag = \sigma_\pm,
	\quad \mu_\pm^\dag = \pm\mu_\pm.
\eeq

Providing the explicit matrix elements for the fields $\sigma_\pm$ and $\mu_\pm$, as above, is of course another way of uniquely defining these fields. The normalizations of the form factors \eqref{ffvac} for the fields $\sigma_-$ and $\mu_-$ with respect to that for the fields $\sigma_+$ and $\mu_+$ are in agreement with \eqref{sZ}. In particular, there are two natural ways of defining $\mu_-$, with a factor $Z$ on the right or on the left of $\mu_+$, differing by a sign. The choice made in \eqref{sZ} implies that $\bra\vac|\mu_-(x)\psi(x')|\vac\ket = -\bra\vac|\mu_+(x)\psi(x')|\vac\ket$, and agreeing with the second equation of \eqref{ffvac}.

\begin{rema} \label{remaexp}
We remark that form factors of the twist fields $\sigma_\pm$ may equivalently be evaluated using Wick's theorem on the particles, with a contraction given by the two-particle form factor. This is an indication that $\sigma_\pm$ are normalized exponentials of bilinear expressions in fermion operators (the overall normalization is made finite by normal ordering). The twist conditions (\ref{twist}) along with the requirement that $\sigma_\pm$ be exponentials of bilinear expressions in the operators $a(\theta)$ and $a^\dag(\theta)$, is a third way to uniquely define, up to normalization, the fields $\sigma_\pm$. Something similar holds for the disorder fields $\mu_\pm$.
\end{rema}

\section{Liouville space and mixed-state form factors}

One usually describes mixed-state averages $\bra\cdots\ket_\rho$ via a trace expression involving the density matrix $\rho$:
\[
	\bra\cdots \ket_\rho := \frc{\Tr\lt(\rho\,\cdots\rt)}{\Tr\lt(\rho\rt)}.
\]
For instance, the (un-normalized) density matrix representing a Gibbs ensemble at a nonzero temperature is $\rho = \rho_\beta:=e^{-\beta H}$, where $\beta$ is the inverse temperature, and with a chemical potential there is the extra factor $e^{-\mu\beta\int_{-\infty}^\infty d\theta\,a^\dag(\theta)a(\theta)}$. The density matrix representing a non-equilibrium steady state sustaining a constant energy flow is \cite{Aschbacher2} (a similar form has been shown in general massice QFT \cite{Ben2}, and see also \cite{DeLuca})
\beq\label{rhoness}
	\rho = \rho_{\rm ness}:=e^{-\beta_l \int_0^\infty d\theta \,E_\theta\, a^\dag(\theta) a(\theta)
	-\beta_r \int_{-\infty}^0 d\theta \,E_\theta\, a^\dag(\theta) a(\theta)}
\eeq
where $\beta_l^{-1}$ and $\beta_r^{-1}$ are the left- and right-temperatures of the asymptotic baths driving the steady state. Further, it has been argued that after quantum quenches, the density matrix becomes the exponential of a linear combination of local conserved charges (generalized Gibbs ensemble)  \cite{Rigol,Rigol2}, and this has been shown in the Ising model \cite{Fagotti,Fagotti1}. Since it is well known that local conserved charges in the Ising model have the form $ \int d\theta\,e^{s\theta} \,a^\dag(\theta)a(\theta)$, also in this case $\rho$ is the exponential of an integral over particle densities $a^\dag(\theta) a(\theta)$.

However, the trace expression of mixed state makes the evaluation of correlation functions (for instance of twist fields of the Ising model) much more involved than in the vacuum. In order to get the full temperature dependence, one needs to re-sum many states. Further, summing explicitly over diagonal matrix elements leads generically to problems of divergency at colliding rapidities. These two problems can be solved by making use of the Liouville space, and the mixed-state form factors.

\subsection{Liouville space}

Note that the main elements necessary to obtain the large-distance form factor expansion (\ref{ff0exp}) are the completeness of the basis of asymptotic states and the fact that the average is evaluated in the vacuum. Thanks to the Gelfand-Naimark-Segal (GNS) construction (see for instance the book \cite{Arveson}), associated to a density matrix (or more precisely to a state, seen as a linear functional on a $C^*$-algebra) there is a vacuum above which one can construct a Hilbert space. The resulting Hilbert space is basically the space of operators (more precisely, a certain completion of a certain quotient of the $C^*$-algebra), and averages with respect to the density matrix are vacuum expectation values in that new Hilbert space. This vacuum expectation value can be expanded using the resolution of the identity, so a ``form factor expansion'' can naturally be obtained for mixed-state two-point functions.

We use these basic ideas here: we consider a space of operators on ${\cal H}$, which we will referred to as the Liouville space (sometimes referred to as the associated Hilbert space) \cite{Fano1,Crawford,Fano2}, and we construct an inner product such that $\bra\cdot\ket_\rho$ becomes a vacuum expectation value in the Liouville space. This will allow us to conjecture a full form factor expansion for mixed-state two-point functions. We will not go into the delicate details of how to actually construct a $C^*$-algebra in order to mathematically have the ingredients necessary for the application of the GNS construction. We will rather provide physical explanations for some of the subtleties involved in the process of obtaining a form factor expansion. These simple Liouville-space ideas are at the basis of the theory of thermofield dynamics (see for instance \cite{Umezawa,Arimitsu}). To our knowledge, the first time form factors were considered in the Liouville space is in \cite{Ben1,Ben3}, for the Ising thermal Gibbs state.

We consider the space of operators with basis formed by $a^{\ep_1}(\theta_1)\cdots a^{\ep_N}(\theta_N)$ for $\theta_1>\ldots>\theta_N$, $\ep_j=\pm$ and $N\in\N$. Here and below, we use
\[
	a^+(\theta):=a^\dag(\theta),\quad a^-(\theta) := a(\theta).
\]
We extend this space to include local fields like $\sigma(x)$ and $\mu(x)$, which can be seen as infinite linear combination (omitting questions completion and of convergence). We denote this by $\End({\cal H})$. The Liouville space $\liou_\rho$ is the inner-product space based on $\End({\cal H})$, with inner product specified by the density matrix $\rho$. With $A,B\in\End({\cal H})$, we denote the corresponding Liouville states by $|A\ket^\rho,\,|B\ket^\rho$ respectively, and we set the inner product to be
\beq\label{ip}
	{}^{\rho}\bra A| B\ket^\rho = \frc{\Tr\lt(\rho\,A^\dag B\rt)}{
	\Tr\lt(\rho\rt)}.
\eeq
We restrict ourselves to density matrices $\rho$ that are diagonal on the asymptotic state basis:
\beq\label{formrho}
	\rho = \exp\lt[-\int d\theta \,W(\theta) \,a^\dag(\theta) a(\theta)\rt].
\eeq
These choices of density matrix include the usual Gibbs state at finite temperature or chemical potential, as well as the non-equilibrium steady state (\ref{rhoness}) and the generalized Gibbs ensembles. 
 
The function $W(\theta)$ should ensure that the result is a well-defined density matrix. We will consider two cases: the {\em untwisted} and the {\em twisted} cases. Let $V(\theta)$ be a function of rapidity that is integrable on the real line, and which we take uniformly positive:
\beq\label{spaceV}
	\inf_{\theta\in\R} V(\theta)>0.
\eeq
The untwisted or twisted cases are, respectively, the choices
\beq\label{WV}
	W(\theta)=V(\theta)
	\quad\mbox{or}\quad W(\theta) = i\pi + V(\theta).
\eeq
In the twisted case, the density matrix $\rho$, on asymptotic states, is a real exponential up to a factor the unitary operator $Z$ \eqref{Z} that implements the $\Z_2$ symmetry (\ref{Z2}).

For convenience, we choose basis elements in the Liouville space formed out of the usual annihilation and creation operators, but with a particular normalization:
\beq\label{basis}
	|\vac\ket^\rho \equiv {\bf 1},\quad
	|\theta_1,\ldots,\theta_N\ket_{\ep_1,\ldots,\ep_N}^\rho
	\equiv Q_{\ep_1,\ldots,\ep_N}^\rho(\theta_1,\ldots,\theta_N)\,a^{\ep_1}(\theta_1)\cdots a^{\ep_N}(\theta_N),
\eeq
where the normalization factors are simply related to the Fermi filling fractions,
\beq\label{Q}
	Q_{\ep_1,\ldots,\ep_N}^\rho(\theta_1,\ldots,\theta_N)
	:= \prod_{i=1}^N \lt(1+e^{-\ep_i W(\theta_i)}\rt).
\eeq
This choice leads to nice analytic properties as will be clear below. In order not to overcount basis elements, we need an ordering, for instance $\theta_1>\ldots>\theta_N$.

Such a structure is sometimes referred to as a ``Liouville-Fock'' space \cite{Schmutz}, and we will refer to a doublet $(\theta,\ep)$ as representing a ``Liouville particle'' of rapidity $\theta$ and type $\ep$. There is of course a hermitian structure inferred from the inner product (the hermitian conjugation still being denoted by ${}^\dag$). One may evaluate the inner products of basis states using the cylic property of the trace and the canonical algebra (\ref{can}):
\beq\label{overlap}
	{}^{\hspace{0.87cm} \rho}_{\ep_1,\ldots,\ep_N}\bra\theta_1,\ldots,\theta_N|\theta_1',\ldots,\theta_N'\ket_{\ep_1',\ldots,\ep_N'}^\rho
	= \prod_{i=1}^N \lt[\lt(1+e^{-\ep_i W(\theta_i)}\rt)
	\delta_{\ep_i,\ep_i'}\delta(\theta_i-\theta_i')\rt]
\eeq
where we assume the ordering $\theta_1>\cdots>\theta_N$ and $\theta_1'>\cdots>\theta_N'$ (note in particular that the states are not ``canonically'' normalized). Let $\la_\pm(\theta)$ and its hermitian conjugate $\la_\pm^\dag(\theta)$ be operators on $\liou_\rho$ (sometimes referred to as ``superoperators'') satisfying the anti-commutation relations
\beq\label{aliou}
	\{\la_\ep(\theta),\la_{\ep'}(\theta')\} =\{\la_\ep^\dag(\theta),\la_{\ep'}^\dag(\theta')\} = 0,\quad \{\la_\ep(\theta),\la_{\ep'}^\dag(\theta')\} = \lt(1+e^{-\ep W(\theta)}\rt)\delta_{\ep,\ep'}\delta(\theta-\theta').
\eeq
The space $\liou_\rho$ can be identified with the Fock space over this algebra,
\beq\label{aliou2}
	\la_\ep(\theta)|\vac\ket^\rho =0,\quad
	|\theta_1,\ldots,\theta_N\ket_{\ep_1,\ldots,\ep_N}^\rho =
	\la_{\ep_1}^\dag(\theta_1)\cdots \la_{\ep_N}^\dag(\theta_N)
	|\vac\ket^\rho.
\eeq

The choice of the ordering $\theta_1>\ldots>\theta_N$ for describing the basis is important, for instance, in the resolution of the identity in terms of basis states (as is apparent for instance in the expansion \eqref{ff0exp} of vacuum correlation functions). It will be convenient however to define $|\theta_1,\ldots,\theta_N\ket_{\ep_1,\ldots,\ep_N}^\rho$ for every $\theta_j\in\R$ by the fact that a sign $(-1)$ is obtained upon exchange of two Liouville particles $(\theta_i,\ep_i)$ and $(\theta_j,\ep_j)$, in agreement with \eqref{aliou}. Then the resolution of the identity can be ``symmetrized'', and re-expressed through integrals over the full line,
\beq\label{res1}
	{\bf 1} = \sum_{N=0}^\infty \sum_{\ep_1,\ldots,\ep_N}
	\int_{-\infty}^\infty \frc{d\theta_1\cdots d\theta_N}{N!\,
	{\prod_{j=1}^{N}\lt(1+e^{-\ep_j W(\theta_j)}\rt)}}
	|\theta_1,\ldots,\theta_N\ket^\rho_{\ep_1,\ldots,\ep_N}
	\;{}_{\ep_1,\ldots,\ep_N}^{\hspace{9mm} \rho}\bra\theta_1,
	\ldots,\theta_N|.
\eeq

See Appendix \ref{appLiou} for more on the formal structure of $\liou_\rho$.

\begin{rema}
We remark that in order for the symmetrized decomposition of the identity to be equivalent to the one where rapidities are ordered, it is crucial to define exchange of Liouville particles as not giving rise to any delta-function ``contact'' term, again in agreement with the first two relations of \eqref{aliou}. By contrast, such contact terms arise upon changing the order of operators $a^+(\theta)$ and $a^-(\theta')$ thanks to the anti-commutation relations (\ref{can}). The Liouville space is formed by a continuous basis without discrete (or delta-function) part at colliding rapidities, so that the colliding-rapidity submanifold of $\R^N$ has zero measure in a decomposition of the identity. There is a {\em caviat} to this in the case of the twist fields, to which we will come back.
\end{rema}

\begin{rema}
It may be necessary for convergence of our form factor expansion below to impose that the real part of $W(\theta)$ grows fast enough at large $|{\rm Re}(\theta)|$. We provide below some indications concerning this.
\end{rema}

\subsection{Definition of mixed-state form factors}

To every operator $A\in\End({\cal H})$ one can associate the Liouville left-action
\beq\label{lefta}
	A^\ell\in \End(\liou_\rho) \;:\; A^\ell|B\ket^\rho = |A B\ket^\rho.
\eeq
In particular, the left-action linear map $A\mapsto A^\ell$ is an algebra homomorphism, $(AB)^\ell = A^\ell B^\ell$. Evidently, averages in the density matrix $\rho$ are then vacuum expectation values in $\liou_\rho$:
\beq
	\bra A\ket_\rho = {}^\rho\bra\vac|A^\ell |\vac\ket^\rho.
\eeq
Hence, using (\ref{res1}), two-point functions should have a spectral decomposition on $\liou_\rho$ where left-action matrix elements are involved. This justifies the following definition of mixed-state form factors associated to $\rho$ (generalizing \cite{Ben1,Ben3}): they are the matrix elements
\beq\label{ff}
	f^{\rho;\Or}_{\ep_1,\ldots,\ep_N}(\theta_1,\ldots,\theta_N)
	:= {}^\rho
	\bra\vac|\Or^\ell|\theta_1,\ldots,\theta_N\ket_{\ep_1,\ldots,\ep_N}^\rho
\eeq
(or more precisely, appropriate analytic extensions thereof),
where $\Or$ is implicitly at the space-time point $(0,0)$. These form factors satisfy the relation
\beq\label{exch}
	f^{\rho;\Or}_{\ep_1,\ldots,\ep_N}(\theta_1,\ldots,\theta_j,\theta_{j+1},\cdots,\theta_N) = -f^{\rho;\Or}_{\ep_1,\ldots,\ep_N}(\theta_1,\ldots,\theta_{j+1},\theta_{j},\cdots,\theta_N)
\eeq
and the cyclic property of the trace implies
\beq\label{otherside}
	{}^{\hspace{0.87cm} \rho}_{\ep_1,\ldots,\ep_N}\bra\theta_1,\ldots,\theta_N|\Or|\vac\ket^\rho =
	f^{\rho;\Or}_{-\ep_N,\ldots,-\ep_1}(\theta_N,\ldots,\theta_1).
\eeq

From this definition, we see that mixed-state form factors are traces with insertions of operators $a^\ep(\theta)$, up to the overall factor $Q_{\ep_1,\ldots,\ep_N}(\theta_1,\ldots,\theta_N)$, and up to the subtraction of contact terms at colliding rapidities, which are absent as explained in the previous subsection. Let us make this more explicit. Traces of products of creation and annihilation operators are evaluated by using Wick's theorem (this follows from cyclicity of the trace and the canonical algebra (\ref{can})) with contractions given by
\[
	Q_{\ep_1,\ep_2}(\theta_1,\theta_2)\bra a^{\ep_1}(\theta_1) a^{\ep_2}(\theta_2)\ket_\rho
	= {}_{-\ep_2}^{\hspace{3mm}\rho}\bra\theta_2|\theta_1\ket_{\ep_1}^\rho.
\]
Traces with a further insertion of a local field, expressed through creation and annihilation operators, are evaluated similarly. This leads in a standard way to a diagrammatic expression, associating a single vertex to the local field. Then, mixed-state form factors are obtained by summing over connected diagrams,
\[
	f^{\rho;\Or}_{\ep_1,\ldots,\ep_N}(\theta_1,\ldots,\theta_N)
	= \Big[Q_{\ep_1,\ldots,\ep_N}(\theta_1,\ldots,\theta_N)\bra
	\Or\, a^{\ep_1}(\theta_1)\cdots a^{\ep_N}(\theta_N)\ket_\rho\Big]_{\rm connected}.
\]
For instance,
\beq\label{Off}
	f^{\rho;\Or}_{\ep_1,\ep_2}(\theta_1,\theta_2) =
	Q_{\ep_1,\ep_2}(\theta_1,\theta_2)
	\bra \Or \,a^{\ep_1}(\theta_1) a^{\ep_2}(\theta_2)\ket_\rho
	- \bra\Or\ket_\rho\;
	{}_{-\ep_2}^{\hspace{3mm}\rho}\bra\theta_2|\theta_1\ket_{\ep_1}.
\eeq

Note that using cyclicity of the trace, the latter can be re-written as
\beq\label{otherside2}
	{}_{\ep_2}^{\hspace{1mm}\rho}\bra\theta_2|\Or^\ell|\theta_1\ket_{\ep_1}^\rho
	=
	f_{\ep_1,-\ep_2}^{\rho;\Or}(\theta_1,\theta_2)+
	\bra\Or\ket_\rho\;
	{}_{\ep_2}^{\hspace{1mm}\rho}\bra\theta_2|\theta_1\ket_{\ep_1}^\rho.
\eeq
Similar equations generalize (\ref{otherside}) and (\ref{otherside2}) to many particles both on the left and on the right.

\begin{rema}
A natural physical interpretation of the Liouville space is as the space of particles and holes excitations created from the Liouville vacuum consisting of a finite density of particles with statistical distribution determined by $\rho$. In this sense, a basis element $|\theta_1,\ldots,\theta_N\ket_{\ep_1,\ldots,\ep_N}^\rho$ represents the presence of $N$ Liouville particles, i.e.~particles ($\ep_j=+$) or holes ($\ep_j=-$) above the finite density. A similar construction to that above can in principle be done for general (at least integrable) QFT, where these particles and holes can scatter. We do not know yet of a clear scattering theory generalizing the usual one to the Liouville space. However, like for the usual pure-state case, we may assume that the form factors (\ref{ff}), as analytic functions of the rapidities continued from the region $\theta>\ldots>\theta_N$, will allow us to extract this scattering matrix: the analytic continuation to different orderings should give rise to particle-exchange properties. We will see below that the choice of a minus sign under permutation of Liouville particles in the vectors $|\theta_1,\ldots,\theta_N\ket_{\ep_1,\ldots,\ep_N}^\rho$ is in agreement with the analytic structure of mixed state form factors for the Ising model: the Ising Liouville scattering theory is, expectedly, still that of free fermions (but with twice as many particles).
\end{rema}

\section{Main results}

\subsection{Mixed-state form factors of local fermion multilinears}

Consider local fields $\Or(x)$ that are normal-ordered products of finitely many fermion fields $\psi(x)$, $\b\psi(x)$ and of their derivatives at a point $x$ (local fermion multilinears). It is a simple matter to evaluate the traces defining mixed-state form factors (or to evaluate any correlation functions) for such fields $\Or$ by using Wick's theorem. However, it is instructive to describe the structure of their mixed-state form factors, which involves a {\em mixing} phenomenon (see Subsection \ref{ssectproof} for proofs of the statements below).

The main statement is that mixed-state form factors of a local fermion multilinear are equal to ordinary matrix elements on ${\cal H}$ of a sum of local fermion multilinears, composed of the original one and of terms with lower numbers of fermions. For instance, one can check explicitly that for the field $\psi$ we have
\[
	f^{\rho;\psi}_\ep(\theta) =\lt\{
	\ba{ll} \bra\vac|\psi|\theta\ket & (\ep=+) \\
	\bra\theta|\psi|\vac\ket & (\ep=-)
	\ea \rt.
\]
and that for the field $\varep$ we have
\[
	f^{\rho;\varep}_{-}(-)  =\frc{m}\pi \int_0^\infty d\theta \frc1{1+e^{W(\theta)}},\quad
	f^{\rho;\varep}_{\ep_1,\ep_2}(\theta_1,\theta_2)=
	\lt\{ \ba{ll}
	\bra\vac|\varep|\theta_1,\theta_2\ket & (\ep_1=\ep_2=+) \\
	\bra \theta_2|\varep|\theta_1\ket & (\ep_1=+,\;\ep_2=-) \\
	\bra \theta_2,\theta_1|\varep|\vac\ket & (\ep_1=\ep_2=-)
	\ea\rt.
\]
where the zero-particle form factor $f^{\rho;\varep}_{-}(-)$ of the field $\varep$ is just its expectation value $\bra\varep\ket_\rho$ (all other form factors are zero). These formulas mean that the mixed-state form factors of $\psi$ and of $\varep$ are equal to matrix elements on ${\cal H}$ of the fields
\beq
	\frak{U}(\psi):=\psi,\quad
	\frak{U}(\varep) := \varep + \bra\varep\ket_\rho {\bf 1}
\eeq
respectively, where the matrix element taken depends on the Liouville particle types. This phenomenon of mixing is in fact completely general: a local field is transformed (or mixed) into a linear combination involving its ``ascendants'' under the free fermion algebra.

More formally, there is a mixing map $\frak{U}$ on the space of local fields such that the following relation holds:
\beq\label{tilde-mapt}
	f^{\rho;\Or}_{{+,\ldots,+,-,\ldots,-}}(\theta_1,\ldots,\theta_j,\theta_{j+1},\ldots,\theta_N)
	= \bra \theta_N,\ldots,\theta_{j+1}|\frak{U}(\Or)|\theta_1,\ldots,\theta_j\ket
\eeq
where there are $j$ plus signs and $N-j$ minus signs as indices on the left-hand side.

One consequence of (\ref{tilde-mapt}) is that mixed-state form factors of normal-ordered product of fermions and their derivatives are entire functions of the rapidities, no matter the analytic properties of $W(\theta)$. This is the reason for our choice of the normalization factor (\ref{Q}).

The map $\frak{U}$ implements the additional ``internal'' contractions present in evaluating the trace that are not present in evaluating usual matrix elements of normal-ordered products. It can be represented conveniently using the Liouville space formalism. Indeed, given a field $\Or$, we can think of the corresponding Liouville state $|\Or\ket^\rho$. Then the map $\frak{U}$ can be represented by the action of an operator on the Liouvile space:
\[
	|\frak{U}(\Or)\ket^\rho = {\lU}|\Or\ket^\rho.
\]
It turns out that the operator $\lU$ performing the necessary Wick contractions is simply
\[
	{\lU} = \exp\lt[\int d\theta\,\frc{{\la}_-(\theta){\la}_+(\theta)}{1+e^{W(\theta)}}\rt].
\]
We may also see these additional contractions as occurring as a consequence of a change of normal ordering, from one with respect to the usual vacuum $|\vac\ket$ to one with respect to the Liouville vacuum $|\vac\ket^\rho$. If we denote by $\nl\cdot \nl$ the normal ordering with respect to $|\vac\ket^\rho$, then we have
\[
	|\Or\ket^\rho = \nl \frak{U}(\Or)^\ell\nl|\vac\ket^\rho.
\]

\begin{rema}
In fact, all fields $\frak{U}(\Or)$ may also be evaluated as follows. One first evaluates them for all fermion linears and bilinears $\Or$; only fermion bilinears get extra contributions, proportional to the identity (one internal contraction). Then, the map $\frak{U}$ on all normal-ordered fields with higher number of fermions can be evaluated recursively by using
\beq\label{cpor}
	[\psi(x),\frak{U}(\Or)] = \frak{U}\lt([\psi(x),\Or]\rt),\quad
	[\b\psi(x),\frak{U}(\Or)] = \frak{U}\lt([\b\psi(x),\Or]\rt),
\eeq
where $[\cdot,\cdot]$ is either the commutator ($\Or$ bosonic) or the anti-commutator ($\Or$ fermionic).
\end{rema}

\begin{rema}
We would like to mention that the use of the map $\frak{U}$ in order to evaluate mixed-state form factors is in parallel with the exponential conformal change of coordinates in conformal field theory, which allows finite-temperature correlation functions to be computed from zero-temperature correlation functions. But this is here in a massive model, and with general density matrices of the form (\ref{formrho}). It would be interesting to gain a fuller geometric or algebraic understanding of it.
\end{rema}

\subsection{Mixed-state form factors of twist fields}

The map $\frak{U}$ in principle allows us to calculate mixed-state form factors from the known matrix elements on ${\cal H}$. However, for the twist fields $\sigma$ and $\mu$, this requires an infinite re-summation: twist fields are infinite linear combinations of normal-ordered products (since they have nonzero matrix elements for arbitrary large number of particles), hence there are infinitely many internal contractions. This re-summation in principle gives rise to two effects: first, the overall normalization of form factors, encoded into the expectation value $\bra\sigma_\pm\ket_\rho = {}^\rho\bra\vac|\sigma_\pm^\ell|\vac\ket^\rho$, is modified from its vacuum value $\bra\sigma\ket$; second, the dependence on the rapidities $\theta_j$ of the form factors is affected. Here we will only consider the dependence on the rapidities.

Using techniques of functional differential equations, we perform ``automatically'' the infinite resummation of contractions in order to obtain the exact rapidity dependence of twist-field form factors. We show in Subsection \ref{ssecttwist} that (here an below, multiple sign possibilities are correlated in any given equation, unless otherwise stated):
\begin{propo} \label{conj}{\ }

\noindent {\bf I.} The one- and two-particle mixed-state form factors of disorder and order fields are given, respectively, by
\beqa
	f^{\rho;\mu_\pm}_{\ep}(\theta)
	&=& \frc1{\sqrt{i}} h_{\ep}^\pm(\theta) \,\bra\sigma_\pm\ket_\rho \n 
	f^{\rho;\sigma_\pm}_{\ep_1,\ep_2}(\theta_1,\theta_2)
	&=& h_{\ep_1}^\pm(\theta_1)h_{\ep_2}^\pm(\theta_2) \lt(\tanh\frac{\theta_2-\theta_1\pm i(\ep_2-\ep_1){\bf 0}}{2}\rt)^{\ep_1\ep_2}\,\bra\sigma_\pm\ket_\rho \label{cff}
\eeqa
where
\beq\label{ch}
	h_\ep^+(\theta) = \sqrt{\frc{i}{2\pi}}\,e^{\ep\frac{ i\pi}4}  \,g(\theta+\ep i{\bf 0})^\ep,\quad
	g(\theta) =  \exp\left[
	\int_{-\infty}^{\infty}
	\frac{d\theta'}{2\pi i}
	\frac1{\sinh(\theta-\theta')}
	\log\left(
	\tanh \frc{W(\theta')}2
	\right)
	\right]
\eeq
and
\beq\label{ch2}
	h_\ep^-(\theta) = \frc1i h_{-\ep}^+(\theta).
\eeq
Here ${\bf 0}>0$ is infinitesimally close to $0$, and the above defines form factors for real rapidities, in general, as distributions obtained from boundary values of analytic functions.

\noindent {\bf II.}  Form factors with higher numbers of particles can be evaluated by using Wick's theorem on the particles. The overall normalization is $\bra\sigma_\pm\ket_\rho$, the contraction of two particles $(\theta_1,\ep_1)$ and $(\theta_2,\ep_2)$ is given by the normalized two-particle form factor $f^{\rho;\sigma_\pm}_{\ep_1,\ep_2}(\theta_1,\theta_2)/\bra\sigma_\pm\ket_\rho$, and the remaining single particle $(\theta,\ep)$, if any, gives a factor $f^{\rho;\mu_\pm}_\ep(\theta)/\bra\sigma_\pm\ket_\rho$; further, there is a minus sign for every crossing of contractions

\noindent {\bf III.} Form factors of $\sigma_\pm$ and $\mu_\pm$ are analytic functions, except for possible ``kinematic poles'' at equal rapidities, in the strip determined by ${\rm Im}(\theta_j)\in(0,\pi)$ for $\ep_j=\pm$ and ${\rm Im}(\theta_j)\in(-\pi,0)$ for $\ep_j=\mp$. Further, for $\theta\in\R$ the functions $h_\ep^\pm(\theta)$ are ordinary integrable functions obtained by continuous continuation from these analyticity regions.
\end{propo}

\begin{rema}
The fact that Wick's theorem can be used to obtain higher-particle form factors is a consequence of the exponential form of twist fields, see Remark \ref{remaexp}.
\end{rema}

\begin{rema}
Mixed-state form factors have the structure of standard vacuum form factors times ``leg-factors'' $h_\ep^\pm(\theta)$ associated to the individual particles $(\theta_j,\ep_j)$. This fact is a consequence of the symmetries of the Ising model (although we will not make use of this fact in our proof). This was first observed in \cite{Fonseca2} in the case of Ising form factors on the cylinder, but the symmetry arguments are based on properties of local fields and on simple dynamical properties of the states, hence they hold as well for form factors on the line in any mixed state considered here. From this viewpoint, the only nontrivial part of our proposition is the expression for the leg factors (\ref{ch}) and (\ref{ch2}).
\end{rema}

\begin{rema}
The positivity condition \eqref{spaceV} guarantees that the integral in \eqref{ch} is well defined. This strong positivity condition is not {\em necessary}, weaker conditions are possible, although this is beyond the scope of this paper.
\end{rema}

One can check using complex conjugation and (\ref{herm}) that the above expressions are in agreement with (\ref{otherside}). Further, the function $h_\ep^\pm(\theta)$ may be analytically continued from the strip
\beq\label{strip}
	I_\ep^\pm:=\lt\{\theta\in\C:\ba{ll}
	{\rm Im}(\theta)\in(0,\pi)& (\ep=\pm)\\
	{\rm Im}(\theta)\in(-\pi,0) & (\ep=\mp)
	\ea\rt\},
\eeq
where it is analytic, to an extended region by extending on both sides of the strip. The extended region depends on the properties of $W(\theta)$ around the real line. Let us assume that $W(\theta)$ is analytic on a neighborhood of some parts of the real line. If $\theta$ lies in this region, and either $\theta$ or $\theta\pm\ep i\pi $ lies in $I_\ep^\pm$, then the analytic continuation is obtained from
\beq\label{feq}
	h_\ep^\pm(\theta) h_\ep^\pm(\theta\pm\ep i\pi)
	= -\frc{\ep}{2\pi}
	\coth \frc{W(\theta)}2.
\eeq

Also, leg-factors with different values of $\ep$ are related to each other:
\beq\label{hphm}
	h_+^\pm(\theta)h_-^\pm(\theta) = \pm \frc{i}{2\pi}
	\coth \frc{W(\theta)}2,
\eeq
this being valid for all $\theta\in \R$ in addition to all values of $\theta$ in the analyticity region of $W(\theta)$. This along with (\ref{feq}) implies that
\beq\label{pluspi}
	h^\pm_-(\theta) = \mp ih_+^\pm(\theta\pm i\pi)
\eeq
whenever the arguments lie in the analytic region of the leg factors. In turn, this implies
\beq\label{shift}
	f^{\rho;\omega_\pm}_{\ep_1,\ep_2,\ldots,\ep_N}
	(\theta_1\pm\ep_1i\pi,\theta_2,\ldots,\theta_N)
	= \pm\ep_1 i\,f^{\rho;\omega_\pm}_{-\ep_1,\ep_2,\ldots,\ep_N}
	(\theta_1,\theta_2,\ldots,\theta_N).
\eeq
This can be seen as a kind of ``crossing symmetry''.

Proposition \ref{conj} was derived in \cite{Ben1,Ben3} in the case of a thermal Gibbs state, both for $W(\theta) = m\beta \cosh\theta$ (untwisted) and $W(\theta) = i\pi + m\beta \cosh\theta$ (twisted), for any $m\beta>0$ (where $\beta$ is the inverse temperature). There, following the lines of the usual vacuum form-factor equations of integrable models, a system of equations and analytic conditions was derived using the underlying free-fermion algebra. Its ``minimal'' solution, assumed to be the correct one for the twist fields $\sigma_\pm$ and $\mu_\pm$, was shown to be (\ref{cff}) (with these $W(\theta)$). General QFT arguments also suggest a relation between nonzero-temperature form factors and form factors on the circle \cite{Ben1}. The latter were independently calculated before \cite{Bugrij,Bugrij3,Fonseca2}, and the minimal solution (\ref{cff}) was shown to be in agreement \cite{Ben3}.

Analyticity requirements, or the relation with form factors on the circle, cannot be used to evaluate mixed-state form factors for general $W(\theta)$. Our proof, in Subsection \ref{ssecttwist}, is based on novel arguments: we derive from the definitions a system of functional differential equations for mixed-state form factors of twist fields as functionals of $W$. Its solution is unique once an initial condition, which can be taken to be at $e^{-W(\theta)} = 0$ (vacuum form factors), has been fixed. We show that the form factors given by Proposition \ref{conj} reproduce the correct vacuum form factors at $e^{-W(\theta)} = 0$, and that they satisfy the system of functional differential equations. In order to gain some intuition into the function (\ref{ch}), we will also provide in Subsection \ref{ssectthermo} ``thermodynamic'' arguments in order to explain the main features of the leg factors $h_\ep^+(\theta)$.

In the case of non-equilibrium steady states, some analyticity conditions can be derived from ``nonequilibrium KMS relation'' for mixed-state form factors, see Subsection \ref{ssectness}. We in Subsection \ref{ssectKMS} show that the analytical properties of form factors from Conjecture \ref{conj} agree with these nonequilibrium KMS relation.

\begin{rema}
Our novel methods provide an alternative proof of the known expression for thermal form factors (i.e. with $W(\theta) = m\beta \cosh\theta$ or $W(\theta) = i\pi + m\beta \cosh\theta$), which, contrary to analytic-property methods, does not require ``minimality'' assumptions.
\end{rema}

\begin{rema}
Twist fields are exponentials of bilinear expressions in fermion operators, see Remark \ref{remaexp}. One may wonder if similar simple expressions for mixed-state form factors hold for general exponential of bilinears. It turns out that the functional-differential equations method is much harder to apply in these cases. However, we derive exact integral-operator expressions for mixed-state form factors of general exponential of bilinear expressions in Subsection \ref{ssectint}.
\end{rema}

\subsection{Mixed-state correlation functions of local fermion multilinears}\label{ssectmultilinears}

The resolution of the identity on the Liouville space, expressed in (\ref{res1}), can be used in order to obtain a series expression for two-point functions in terms of form factors (\ref{ff}), leading to an expansion similar to (\ref{ff0exp}). Taking into account the state normalization (\ref{overlap}), the resolution of the identity gives
\beq\label{ffexp}
\bra \Or(x,t)\Or^\dag(0,0)\ket_\rho
= \sum_{N=0}^\infty \sum_{\ep_1,\ldots,\ep_N}
	\int_{-\infty}^\infty \frc{d\theta_1\cdots d\theta_N}{N!}\frac{e^{\sum_{j=1}^N \lt(i\ep_jp_{\theta_j} x - i\ep_jE_{\theta_j} t\rt)}}{\prod_{j=1}^{N}\lt(1+e^{-\ep_j W(\theta_j)}\rt)}\lt| f^{\rho;\Or}_{\ep_1,\ldots,\ep_N}
	(\theta_1,\ldots,\theta_N)\rt|^2.
\eeq
Whenever $\Or$ is a field whose form factors are zero for large enough numbers of particles, in which case the above series truncates, then we expect this to be a correct expression. Indeed, we expect that in these cases, the full GNS construction can be perform within an appropriate free-fermion framework. Local fermion multilinears, for instance the fermion fields $\psi$ and $\b\psi$ or the field $\varep$, are such examples of $\Or$.

The integrals in \eqref{ffexp}, however, require some analysis. As we noted, form factors of fermion multilinears are entire functions of the rapidities. If $W(\theta)$ has increasing real part as $|\theta|\to\infty$, then the integral over $\theta_j$ is convergent if $\ep_j=-$. However, the integral over $\theta_j$ is in general not convergent if $\ep_j=+$. One may make both cases $\ep_j=\pm$ convergent if ${\rm Re}(W(\theta))$ grows like, or faster than, $e^{\alpha\cosh\theta}$ for some $\alpha>0$ as $|{\rm Re}(\theta)|\to\infty$, by adding a small negative imaginary part to $t$. The correlation function is then seen as the boundary value, at $t\in\R$, of a function of $t$ analytic on some neighborhood of $\R$ in the region ${\rm Im}(t)>0$. This is a standard prescription for QFT correlation functions.

This boundary value is finite at space-like distances ($x^2>t^2$) for any $W(\theta)$ that is analytic on neighborhoods of parts of the real line unbounded both towards $\pm\infty$. Indeed, in this case, assuming without loss of generality that $x>0$, one can then shift the contours as $\theta_j \mapsto \theta_j+i\eta\ep_j$ in the region $\lt|{\rm Re}(\theta_j)\rt|>K$, for $K>0$ large enough and $\eta>0$ small enough in such a way that $\theta_j$ remains in the analyticity region of $W(\theta_j)$. In fact, if $W(\theta)$ does not grow like or faster than $e^{\alpha\cosh\theta}$, but is analytic on neighborhoods of parts of the real line unbounded both towards $\pm\infty$, one should {\em define} the correlation function by \eqref{ffexp} with integrals on shifted contours as explained.

\begin{rema} The condition \eqref{spaceV} that the real part of $W(\theta)$ be uniformly positive for real $\theta$ is not necessary here if the imaginary part of $W(\theta)$ is zero (untwisted case). It is necessary if the imaginary part is $i\pi$ (twisted case), in order to avoid the presence of poles in \eqref{ffexp}.
\end{rema}

\subsection{Mixed-state correlation functions of twist fields}

For twist fields $\sigma_\pm$ and $\mu_\pm$, the form factor expansion is infinite, as these fields have non-zero form factors for arbitrary large numbers of particles. However, the resulting infinite series (\ref{ffexp}) is not the correct representation of the two-point function. The form factor expansion is modified in various ways, because of the cuts emanating from the fields $\sigma_\pm$ and $\mu_\pm$ as expressed in the twist condition \eqref{twist}. We provide intuitive arguments for the modifications involved and a conjecture for the exact series expansion. Throughout we take again without loss of generality $x>0$, and we concentrate solely on the space-like region $x^2>t^2$.

\subsubsection{Three modifications}

First, consider the general form of the expansion $\sum_s {}^\rho\bra\vac|\Or(x,t)|s\ket^\rho \;{}^\rho\bra s|\Or^\dag(0,0)|\vac\ket^\rho$. If this is to be a large distance expansion, then we should interpret the intermediate states $|s\ket^\rho\;{}^\rho\bra s|$ as lying, or carrying correlations, over the region between the fields, and the vacuum states ${}^\rho\bra\vac|$ and $|\vac\ket^\rho$ as representing what is happening far on the right and left, respectively. In the regions where there is a cut due to a twist field, the density matrix is effectively locally modified by the unitary operator $Z$ \eqref{Z} associated to the $\Z_2$ transformation:
\beq\label{Wsharp}
	\rho\mapsto \rho^\sharp:=Z\rho:\quad W(\theta)\mapsto W^\sharp(\theta) := W(\theta)+i\pi.
\eeq
It was argued in \cite{Ben3}, in the case of the finite-temperature form factors and via a comparison with form factors on the circle, that the intermediate states must lie in a region that is not affected by the cuts. This is in order for the space of excited states constructed, and the form factors obtained, to correspond to the sector represented by $\rho$, and not the modified one $\rho^\sharp$. Hence, form factor expansions can be obtained in the cases where the cuts emanate {\em away} from the region between 0 and $x$, so that no cut is present {\em between} 0 and $x$:
\beq\label{type}
	\bra \sigma_+(x,t)\sigma_-(0,0)\ket_\rho,\quad
	\bra \mu_+(x,t)\mu_-(0,0)\ket_\rho.
\eeq

On the other hand, according to \eqref{sl}, the scaling limit leads to correlation functions \eqref{GGt}, $G(x,t)=\bra\sigma_+(x,t)\sigma_+(0,0)\ket_\rho$ (ordered regime) or $\t G(x,t)=\bra \mu_+(x,t)\mu_+(0,0)\ket_\rho$ (disordered regime). Here the cuts emanating from each field in the correlation function are going in the same direction (towards the right). We recall that this is because in the Jordan-Wigner transformation, the Pauli spin matrices are written in terms of infinite products of fermion operators starting at the matrix's site and going in a fixed direction (for instance towards the right). As a consequence, there is a cut {\em between} 0 and $x$, and our form factors, according to our arguments above, cannot directly be used.  Happily, we may use \eqref{sZ} to relate $\sigma_+$ and $\mu_+$ to $\sigma_-$ and $\mu_-$ via the unitary operator $Z$ \eqref{Z}, and obtain correlation function of the type \eqref{type}. With \eqref{Wsharp}, we find
\beq\label{chgcut}
	G(x,t)
	= \bra \sigma_+(x,t)\sigma_-(0,0)\ket_{\rho^\sharp},\quad
	\t G(x,t)
	= \bra \mu_+(x,t)\mu_-(0,0)\ket_{\rho^\sharp}.
\eeq
According to our previous discussion, form factor expansions can be obtained for the correlation functions on the right-hand side, so that these are more useful expressions for $G(x,t)$ and $\t G(x,t)$ than the original ones \eqref{GGt}.

Second, the presence of a cut affects the translation covariance property of $x$-dependent matrix elements, providing a real exponential factor. This property can be expressed as
\beq\label{transcov}
	{}^\rho\bra \vac|\omega_\pm(x,t)|\theta_1,\ldots,\theta_N\ket^\rho_{\ep_1,\ldots,\ep_N}
	=
	e^{\pm x{\cal E}}
	e^{\sum_{j=1}^N \lt(i\ep_jp_{\theta_j} x - i\ep_jE_{\theta_j} t\rt)}
	f^{\rho;\omega}_{\ep_1,\ldots,\ep_N}(\theta_1,\ldots,\theta_N)
\eeq
for both order and disorder twist fields $\omega=\sigma$ and $\omega=\mu$, with ${\cal E}$ the {\em free energy deficit}, whose expression is given in \eqref{E}. The factor $e^{\pm x{\cal E}}$ can be understood physically by the fact that the trace in the region where a cut lies contributes a different free energy to that where no cut lies, so that, after appropriate normalization with respect to the point $x=0$, a shift of the cut end-point gives rise to the exponential of the twist-field free energy deficit. We will provide a thermodynamic argument for the expression \eqref{E} in Subsection \ref{ssectthermo}. Denoting ${\cal E}^\sharp$ the free energy deficit associated to $W^\sharp$ (see \eqref{Wsharp}), we have
\beq\label{Esharp}
	{\cal E}^\sharp = - {\cal E}.
\eeq

The final subtlety is that the form factors of twist fields are not entire functions of the rapidities -- they are distributions defined as boundary values of analytic functions, and it appears to be important, in order to obtain a well-defined form factor expansion, to be able to shift contours towards the analyticity regions. Hence, we need to further require that $W(\theta)$ be analytic on a neighborhood of $\R$. This requirement is not satisfied for the non-equilibrium steady state (\ref{rhoness}), and we will analyze the consequence of this in Subsection \ref{ssectness}.

\begin{rema}
It is evident that $\bra\sigma_+(x,t)\sigma_+(0,0)\ket_\rho$ and $\bra \mu_+(x,t)\mu_+(0,0)\ket_\rho$ are real, because by time and space translation invariance, they can be brought to the form of an average of an operator times its hermitian conjugate. However, the expressions on the right-hand side of \eqref{chgcut} are not obviously real.  The verification of the reality of the form factor expansion obtained will be a nontrivial check.
\end{rema}

\subsubsection{The mixed-state form factor expansion}

Putting these subtleties together, we obtain the following.
\begin{conjecture} \label{conjexp}
Both for $\omega=\sigma$ and for $\omega=\mu$, and with $W(\theta)$ analytic on a neighborhood of $\theta\in\R$, we have
\begin{eqnarray} \label{correlation}
\lefteqn{\bra\omega_{+}(x,t) \omega_{+}(0,0)\ket_\rho}\\
&=&
e^{-x{\cal E}}\sum_{N=0}^{\infty} \sum_{\ep_1,\ldots,\ep_N}\int \frac{d\theta_1 \cdots d\theta_N}{N!}\frac{e^{\sum_{j=1}^N \lt(i\ep_jp_{\theta_j} x - i\ep_jE_{\theta_j} t\rt)}}{\prod_{j=1}^{N}\lt(1-e^{-\ep_j W(\theta_j)}\rt)}
f^{\rho^\sharp;\omega_{+}}_{\ep_1,\ldots,\ep_N}(\theta_1,\ldots,\theta_N) f^{\rho^\sharp; \omega_{-}}_{-\ep_N,\ldots,-\ep_1}(\theta_N,\ldots, \theta_1)
\no
\end{eqnarray}
Recall that $\rho^\sharp$ is the density matrix corresponding to $W^\sharp$ in (\ref{Wsharp}).
\end{conjecture}
In order to arrive at this, we first used  (\ref{chgcut}), then from this the expansion obtained from (\ref{res1}), then (\ref{transcov}) and (\ref{otherside}), and finally the $\sharp$ modifications given by (\ref{Esharp}) and (\ref{Wsharp}). Using the mixed-state form factors expressed in Proposition \ref{conj}, and recalling \eqref{GGt}, one can simplify this to \eqref{corrsimple}, where the proportionality constant is the product of expectation values $\bra\sigma_+\ket_{\rho}\bra\sigma_-\ket_{\rho}$. We provide in Appendix \ref{appreal} a check to one-particle order that the result is a real function, as it should be from general arguments. Conjecture (\ref{conjexp}) was shown to be correct for thermal Gibbs states in \cite{Ben3} by showing, through contour shifts, that it reproduces the correct form factor expansion in the quantization on the circle.

As in the case of fermion multilinears explained in Subsection \ref{ssectmultilinears}, in order to obtain a large-distance expansion from the conditionally convergent integrals, we shift the $\theta_j$ contour in (\ref{correlation}) by $i\ep_j \eta$ for $\eta>0$ small enough in such a way that $\theta_j$ remains in the analyticity region of $W(\theta_j)$. Note that these shifts keep the rapidities in the analyticity region of the form factors involved.

Shifting the contours further, one may hit singularities of the function $(1-e^{\pm W(\theta)})^{-1}$. These singularities then determine the large-distance asymptotic behavior of the two-point function. With $\theta^\star$ the minimum of $|{\rm Im}(\sinh\theta)|$ over the singularities in $\theta$ of $\lt(1-e^{\pm W(\theta)}\rt)^{-1}$, one has
\beq\label{leadingnext}
	G(x,0)
	= \bra\sigma_+\ket_{\rho}
	\bra\sigma_-\ket_{\rho}  e^{-x{\cal E}}\lt(1 +
	O\lt(e^{- 2mx |{\rm Im}(\sinh \theta^\star)|}\rt) \rt),\quad
	\t G(x,0) = 
	O\lt(e^{- mx({\cal E}+ |{\rm Im}(\sinh \theta^\star)|)}\rt)
\eeq
Here, the exponential decay includes possible algebraic or other non-exponential factors in $mx$. These factors are determined by the type of singularities. We provide examples of this in our analysis of a particular generalized Gibbs ensemble in Sub-section \ref{ssectquantq}, and in our analysis of the non-equilibrium steady state in Sub-section \ref{ssectness}.

\begin{rema}
As we mentioned, in the case of the non-equilibrium steady state, because of the non-analyticity at $\theta=0$, care must be taken in performing the shifts, and this affects the large-distance behaviour (see Subsection \ref{ssectness}). In particular, although the form of the leading exponential decay $e^{-x{\cal E}}$ is still valid, the subleading terms are expected to be slower than exponential. The factor $e^{-x{\cal E}}$ is in agreement with the result of \cite{Aschbacher} in the anisotropic XY model out of equilibrium.
\end{rema}

\section{Applications}

\subsection{Quantum quenches} \label{ssectquantq}

Quantum quenches are physical setups where, in the usual convention, one suddenly modifies one or more parameters (couplings, etc.) of a system, usually initially in its ground state, and let it evolve. Hence in general, one has two hamiltonians $H$ and $H'$ which do not commute with each other, and the ground state of $H$ (or some low-energy excitation, or some natural mixed state) is unitarily evolved with the evolution operator $e^{-itH'}$. The main problem is to determine what happens after a large evolution time $t$; for instance: is there thermalization? A ground breaking experiment \cite{Kin} showed that dimensionality plays a crucial role, and suggested that conservation laws may have a strong influence on the result. A large body of work has been done in this subject (for a comprehensive summary, see for instance \cite{Pol}). When measuring averages of local operators in infinite volume, one would indeed expect the large-time results to be consistent with thermalization, with a temperature that depends on the actual quench performed and on the initial state. It has been observed that, starting with ground states, thermalization indeed occurs in non-integrable systems, but that the infinite number of conserved quantities in integrable systems restrict the dynamics enough so that usual thermalization does not occur. In integrable systems, the final state is rather one described by a generalized Gibbs ensemble (GGE) \cite{Rigol,Rigol2}. A GGE is expected to have a density matrix of the form
\beq\label{GGE}
        \rho_{\rm GGE} = e^{-\sum_{n=1}^\infty \beta_n H_n}
\eeq
where $H_n$ are the local conserved quantities, and $\beta_n$ are the associated generalized inverse temperatures. The generalized temperatures are expected to be determinable by the requirement that the averages of the conserved densities in the GGE be equal to those in the initial state.

In GGEs, one usually expects only conserved quantities that are bounded from below to appear in \eqref{GGE}, one assumes that the series in the exponential in \eqref{GGE} has convergence properties, and one usually does not expect the state to admit flows of energy, particles, etc. Because of the latter, although the state is not strictly at equilibrium (we do not have a standard Gibbs' ensemble), it is in a natural ``generalized'' equilibrium: there is no entropy production.

In the Ising model, a precise study of quenches shows that indeed the GGE occurs \cite{Fagotti,Fagotti1,Fagotti2}. Our results can be applied directly to GGEs in the Ising model thanks to the trivial observation that in the Ising model, conserved quantities are linear combinations (integrals) of $a^\dag(\theta)a(\theta)$ (a similar statement holds for general integrable QFT, using appropriate asymptotic-state creation and annihilation operators). One may construct generalized hamiltonians $H_n=(Q_n+Q_{-n})/2$ using charges $Q_n$ defined as
\beq
        Q_n = \int d\theta\, e^{n\theta}\,a^\dag(\theta)a(\theta)
\eeq
(and in particular $H_1$ is the usual hamiltonian), and one has, according to (\ref{formrho}),
\beq
        W(\theta) = \sum_{n=1}^\infty \beta_n \cosh(n\theta).
\eeq
But more general functions $W(\theta)$ are allowed by our formalism. Hence our results (\ref{ffexp}) (for fermion fields and all their descendants) and (\ref{correlation}) (for order and disorder fields) give, in principle, the full large-distance expansion of correlation functions in the Ising model in any GGE.

A particular quench corresponding to an abrupt change of the external magnetic field from $h_0$ to $h$ was considered in \cite{Fagotti,Fagotti1,Fagotti2}. In the scaling limit, with both $h_0$ and $h$ near to 1, this corresponds to a change of mass from $m_0$ to $m$. The approach of the order parameter (one-point function), in the ordered phase, to its steady average value after the quench was studied using form factors in \cite{Schuricht2012}, reproducing the results of \cite{Fagotti}. Here we consider instead the two-point function, directly in the steady state. The result of the works \cite{Fagotti,Fagotti1,Fagotti2} can be expressed, in the scaling limit, by the following choice of $W(\theta)$:
\beq\label{WGGE}
	\tanh\frc{W(\theta)}2 = \frc{\sinh^2\theta + \eta \,m_0/m}{\cosh\theta \sqrt{
	\sinh^2\theta + m_0^2/m^2}}
\eeq
where $\eta=+$ for a quench from ferromagnetic to ferromagnetic or from anti-ferromagnetic to anti-ferromagnetic regimes, and $\eta=-$ for the other cases. We see that for $\eta=-$ this is out of the context that we considered, since then $W(\theta)\leq 0$ for small enough values of $\theta$. However for $\eta=+$ this can be treated with the present formalism. In particular, we find that
\beq\label{WU}
	\frc1{1-e^{-W}} = \frc12(1+U),\quad
	\frc1{1-e^W} = \frc12 (1-U)
\eeq
where
\beq\label{UGGE}
	U(\theta):=\frc{\cosh\theta \sqrt{
	\sinh^2\theta + m_0^2/m^2}}{\sinh^2\theta + m_0/m}
\eeq
Hence there are poles at
\[
	\sinh\theta =\pm i\, \sqrt{m_0/m}
\]
and branch points at
\[
	\sinh\theta =\pm i\,m_0/m.
\]
This implies that the expansion (\ref{corrsimple}) provides a full large-distance expansion for correlation functions of ordered and disordered fields in the universal stationary regime occurring after such magnetic-field quenches, and in particular that the large-distance behavior is of the form (\ref{leadingnext}) with exponential decay controlled by
\[
	|{\rm Im}(\sinh\theta^\star)| =  \sqrt{m_0/m}\quad \mbox{or}\quad
	 m_0/m.
\]

We may perform a short analysis in order to make the statement about exponential decay more precise, and show that our results agree with the general features of the results of \cite{Fagotti2}. First, in the ordered (ferromagnetic) regime, we must use the order field $\sigma_+$. According to (\ref{leadingnext}), we then see, at leading order, a pure exponential decay $O\lt(e^{-{\cal E}x}\rt)$. Indeed a pure exponential decay was derived in \cite{Fagotti2}. In the disordered (anti-ferromagnetic) regime, we must use the disorder field $\mu_+$. We now see in (\ref{leadingnext}) an exponential decay, with a possible algebraic factor. In order to evaluate the possible algebraic factor, we simply need to deform the $\theta$ contours, and look at the singularities, at points $z$, making $|{\rm Im}(\sinh z)|$ minimum. If $m_0/m>1$, then $\sqrt{m_0/m} < m_0/m$, hence the poles are these singularities. Poles provide a pure exponential behavior, so we have a pure exponential decay $O\lt(e^{-({\cal E}+\sqrt{mm_0})x}\rt)$. A pure exponential behavior, without algebraic factors, was indeed found in \cite{Fagotti2} for $h_0>h$ in the anti-ferromagnetic regime. If $m_0/m<1$, then $\sqrt{m_0/m} > m_0/m$, and the branch point is the nearest singularity. Hence we have an exponential decay, but branch cuts give additional algebraic factors. Indeed, an algebraic factor is found in \cite{Fagotti2} in this situation. The branch cut is of square-root type, so that, after contour shifts, the algebraic factor is obtained by evaluating
\[
	e^{-m_0x }
	\int_{0}^\infty d\ell\, \sqrt{\ell}\,
	e^{-Cmx\,\ell} \propto (mx)^{-3/2}\,e^{-m_0x} \no
\]
for some $C>0$. Hence in this case, the overall leading behavior is $O\lt(x^{-3/2} e^{-({\cal E} + m_0)x}\rt)$. An algebraic factor with exponent $-3/2$ was indeed found in \cite{Fagotti2}.

We may also describe the correction terms, which were not described in \cite{Fagotti2}. We find for instance, in the ordered regime, considering the two-particle contributions of the form factor expansion, the correlation function $G(x)$ at large $x$ proportional to
\[
	\ba{ll}
	e^{-{\cal E}x}\lt(1+O\lt(e^{-2\sqrt{mm_0}x}\rt)\rt)
	& \mbox{ordered, $m_0>m$} \\
	e^{-{\cal E}x}\lt(1+O\lt(x^{-3}e^{-2m_0x}\rt)\rt)
	& \mbox{ordered, $m_0<m$}.
	\ea
\]
See appendix \ref{appleading} for the some details of these calculations.

Of course, a similar, but simpler, analysis may be done for the field $\varep$, using the finite-sum exact expression (\ref{ffexp}).

Hence, a simple form factor analysis provides the main features of some results of \cite{Fagotti2} as well as some subleading terms. A detailed analysis of the large-distance expansion from our exact result (\ref{corrsimple}), further confirming and generalizing \cite{Fagotti2}, would be very interesting, but is beyond the scope of this paper.

\subsection{Non-equilibrium steady states} \label{ssectness}

A non-equilibrium steady state (NESS) can be seen as a state which is steady with respect to the dynamics of the model, but where there are flows of energy, particles, charge, etc with constant rate. Non-equilibrium steady states can be ``created'' by evolving unitarily, for an infinite time, two semi-infinite halves of a system initially separately thermalized. If the two halves are thermalized at different temperatures $\beta_l^{-1}$ and $\beta_r^{-1}$, then the steady state obtained possesses energy flows. This state was studied in critical systems in \cite{Ben4,BDaihp} using conformal field theory, near to criticality in  \cite{Ben2} using general massive quantum field theory and in \cite{CAetal} using integrability, and in the XY quantum chain (Ising model) in \cite{Aschbacher2}, \cite{DeLuca}. The results for the Ising model can be stated, in our present notation, by saying that the function $W(\theta)$ is specialized to
\beq\label{Wness}
	W_{\rm ness}(\theta) := \beta_l E_\theta \Theta(\theta)
	+\beta_r E_\theta \Theta(-\theta)
\eeq
where $\Theta(x)$ is the Heavyside step function (we denote by $\rho_{\rm ness}$ the associated density matrix). This associates an inverse temperature $\beta_l$ to right-moving particles, and an inverse temperature $\beta_r$ to left-moving ones. Physically, this formula is justified by the fact that right-moving particles come from the far left, which was thermalized at temperature $\beta_l^{-1}$, and vice versa.

\subsubsection{Non-equilibrium form factors have branch cuts}

Since the function $W_{\rm ness}(\theta)$ is analytic on ${\rm Re}(\theta)<0$ and on ${\rm Re}(\theta)>0$ but not at $\theta=0$, it does not satisfy the assumptions that we made for the form factor expansion. This means that we can analytically continue form factors $f^{\rho_{\rm ness};\omega_\pm}_{\ep_1,\ldots,\ep_N}(\theta_1,\ldots,\theta_N)$, for $\omega=\mu$ or $\omega=\sigma$, from their analyticity strip $\theta_j\in I_{\ep_j}^\pm$ (\ref{strip}) whenever ${\rm Re}(\theta)<0$ or ${\rm Re}(\theta)>0$ using (\ref{feq}), but not from the point $\theta=0$. For $\beta_l\neq\beta_r$, we can see from (\ref{feq}) that there is a jump, in this analytic continuation, in going from positive to negative real part of $\theta$. This means that the functions $h_\ep^\pm(\theta)$ (hence also all form factors) have branch points at $\theta=0$ and $\theta \pm\ep i\pi$ with branch cuts running away from the analyticity region. The jump through these branch cuts can be evaluated using (\ref{feq}) and (\ref{Wness}), and is given by
\beq\label{jump}
	\frc{h_\ep^{\pm}(\theta-{\bf 0})}{h_\ep^{\pm}(\theta+{\bf 0})}
	= \lt\{\ba{ll}
	\coth\frc{m\beta_r\cosh\theta}2 \;\tanh\frc{m\beta_l\cosh\theta}2
	&  \mbox{(for $W_{\rm ness}$)}
	\z
	\tanh\frc{m\beta_r\cosh\theta}2 \;\coth\frc{m\beta_l\cosh\theta}2
	& \mbox{(for $W_{\rm ness}^\sharp$)} \ea\rt.
\eeq
for ${\rm Re}(\theta)=0$ and $\theta\pm \ep i\pi\in I_\ep^\pm$. In fact, one can evaluate the leading small-$\theta$ behavior by extracting the pole $1/(\theta-\theta')$ from the factor $1/\sinh(\theta-\theta')$ in (\ref{ch}), thus obtaining
\beq\label{smalltheta}
	h_\ep^\pm(\theta) \propto \lt\{\ba{ll}
	\theta^{\mp i\ep \gamma} &
	 \mbox{(for $W_{\rm ness}$)} \\
	\theta^{\pm i\ep \gamma} &
	 \mbox{(for $W_{\rm ness}^\sharp$)}
	 \ea\rt.
	,\quad \gamma:=\frc1{2\pi}\log\lt(
	\coth\frc{m\beta_r}2 \;\tanh\frc{m\beta_l}2\rt).
\eeq

\subsubsection{Expansion for two-point functions: oscillations in $\log (mx)$}

We can attempt to make the expansion (\ref{correlation}) into a large-distance expansion by shifting contours. This is obtained by shifting the $\theta_j$-contours towards the positive imaginary direction by an amount $+i\pi$, for every rapidity associated with $\ep_j=+$. In shifting the contour, poles are picked up from the factors $1-e^{-\ep_jW_{\rm ness}(\theta_j)}$ in the denominator; and the segment running from 0 to $i\pi$ is surrounded in such a way that the contour passes by 0, because the factors $1-e^{-\ep_jW(\theta_j)}$ are not continuous at $0$. Fully shifted contours with $\ep=+$ cancel unshifted contours with $\ep=-$ thanks to (\ref{shift}), so that one is left with pole contributions and integrals along the segment from 0 to $i\pi$. The result is, formally (see Appendix \ref{appness} for some details),
\beqa
	\lefteqn{\bra\omega_+(x,t)\omega_+(0,0)\ket_{\rho_{\rm ness}}}
	&& \n &=&
	e^{-x{\cal E}_{\rm ness}}
	\sum_{N,M=0}^\infty \sum_{n_1,\ldots,n_N\in\Z\setminus\{0\}}
	\int_{0}^{\pi}
	 \frc{d\theta_1\cdots d\theta_M}{N!M!}\,
	(2\pi)^N\, i^M\times \n && \times\;
	\prod_{j=1}^N \frc{
	e^{-mx\cosh(\alpha(n_j)) + 2\pi n_j t/\beta(n_j)}}{
	m\beta(n_j) \cosh \alpha(n_j)
	}\times \n && \times\;
	\prod_{k=1}^{M} \lt[
	e^{-mx\sin(\theta_k) -imt\cos(\theta_k)} \lt(
	\frc1{1-e^{-m\beta_l\cos(\theta_k-i{\bf 0})}}
	-
	\frc1{1-e^{-m\beta_r\cos(\theta_k+i{\bf 0})}}
	\rt)
	\rt]
	\times \n && \times\;
	F\lt(
	\alpha(n_1)+\frc{i\pi}2,\ldots,\alpha(n_N)+\frc{i\pi}2,
	i\theta_1,\ldots,i\theta_N\rt)
	\label{resness}
\eeqa
where
\beq
	F(u_1,\ldots,u_K) =
	f_{+,\ldots,+}^{\rho_{\rm ness}^\sharp;\omega_+}(u_1,\ldots,u_K)
	f_{-,\ldots,-}^{\rho_{\rm ness}^\sharp;\omega_-}(u_K,\ldots,u_1)
\eeq
and
\beq\label{alphabeta}
	\alpha(n) = {\rm arcsinh}\lt(\frc{2\pi n}{\beta(n)m}\rt),\quad
	\beta(n) = \lt\{\ba{ll} \beta_l & (n>0) \\ \beta_r & (n<0) \ea\rt..
\eeq
Here again, either $\omega=\sigma$ or $\omega=\mu$.

We see that the standard Matsubara frequencies of finite-temperature correlation functions appear on the second line of the right-hand side in (\ref{resness}); however here there are two temperatures, hence two frequencies, depending on the sign of $n_j$. The Matsubara frequencies can be interpreted as coming from the quantization of the momentum in a quantum system on the circle, since a finite-temperature state can be interpreted as a condition of (quasi-)periodicity in imaginary time, with extent given by the inverse temperature. Here, we see that we have the remnant of this in our non-equilibrium steady state, but with two different circumferences, for right- and left-moving particles. Separate (quasi-)periodicity circumferences in imaginary time for right- and left-movers does not make full sense in the massive theory, as right- and left-movers do not separate (fields do not factorize). This is why we have extra integrals on finite intervals; these can be interpreted as providing the bridge for right- and left-moving modes to jump from the circle of circumference $\beta_l$ to that of circumference $\beta_r$ and vice versa.

The leading exponential decay $e^{-x{\cal E}_{\rm ness}}$ is in agreement with the result of \cite{Aschbacher} obtained in the (anisotropic) XY model out of equilibrium, after one performs the scaling limit. The non-equilibrium ``free energy deficit'' ${\cal E}_{\rm ness}$ can be written as the average of the free energy deficits ${\cal E}_\beta$ associated to equilibrium thermal density matrices,
\beq
	{\cal E}_{\rm ness} = \frc12\lt({\cal E}_{\beta_l}+{\cal E}_{\beta_r}\rt),\quad
	{\cal E}_\beta
	:= \int_{-\infty}^\infty \frc{d\theta}{2\pi}
	m\cosh\theta \log\lt(
	\coth\frc{m\beta \cosh(\theta)}2
	\rt).
\eeq

The form of the leading or subleading terms multiplying this exponential decay however is more subtle to obtain. Because of the non-analyticity of $W(\theta)$ at $\theta=0$, it is difficult to analyze the relative strength of the terms with different particle numbers. Under the hypothesis that the one- and two-particle contributions give the correct form, perhaps up to normalizations, we may perform the following analysis.

We consider the one- or two-particle integral and zero-particle sum, $M=1,2$ and $N=0$; these are not exponential, hence we can neglect all exponentially decaying terms coming from higher values of $N$. They are evaluated by approximating the integral over $[0,\pi]$ of $e^{-mx\sin(\theta)} c(\theta)$ by the integral over $[0,\infty)$ of $e^{-mx\theta} (c(\theta)+c(\pi-\theta))$. Let us take $t=0$ for simplicity. Using (\ref{smalltheta}) and (\ref{feq}), and the fact that the two-point function is real, this gives in the case of the disorder field ($M=1$):
\beq\label{expmu}
	\bra\mu_+(x,0)\mu_+(0,0)\ket_{\rho_{\rm ness}}
	= e^{-x{\cal E}_{\rm ness}} \bra\sigma_+\ket_{\rho_{\rm ness}}
	\bra\sigma_-\ket_{\rho_{\rm ness}}\lt(
	\frc{A} {mx} \cos(2\gamma\log(mx)+B)+\ldots\rt)
\eeq
where $\gamma$ is defined in (\ref{smalltheta}).

The constants $A$ and $B$ can also be evaluated from the $M=1$ integral after a somewhat tedious analysis (see Appendix \ref{appness} for the answer). However, we do not expect this $M=1$ evaluation of the constants $A$ and $B$ to be meaningful. Indeed, the omitted part in (\ref{expmu}), corresponding to higher values of $N$ and $M$, contain terms of exactly the same form as the first subleading term written in (\ref{expmu}), but with different constants $A$ and $B$, with $A$ possibly logarithmically divergent. These come from higher values of $M$ still with $N=0$, and arise due to the ``kinematic poles'': the singularities of the $\coth((\theta_1-\theta_2)/2$ factors -- see Appendix \ref{appness} for some details. We do not have a clear interpretation of the potential logarithmic divergence of the constant $A$, but it may correspond to the large-time limit taken to form the steady state (at any finite time after a quench, the density matrix is a smooth version of $W_{\rm ness}(\theta)$).  Yet the oscillating form obtained from the $M=1$ analysis may be correct, and the possible divergences may be re-absorbable into the normalization of the field. Note that other subleading terms are terms with higher oscillating frequency in $\log(mx)$ and with higher power of $1/(mx)$ (coming from higher values of $M$ with $N=0$), as well as exponentially decaying terms.

A similar analysis for the order field is obtained from $M=2$. In this case, the first subleading term is $O(1)$; this is for the same reason as that for which the next subleading term in the disorder-field case are of the same form as the first subleading term. Hence, also in this case, the vacuum expectation value $\bra\sigma_\pm\ket_{\rho_{\rm ness}}$ may not not reproduce the correct normalization for the leading exponential decay of the two-point function, $\bra\sigma_+(x,0)\sigma_+(0,0)\ket_{\rho_{\rm ness}} \sim C \,e^{-x{\cal E}_{\rm ness}}$ with in general $C\neq \bra\sigma_+\ket_{\rho_{\rm ness}}\bra\sigma_-\ket_{\rho_{\rm ness}}$. In this case the full physical meaning, beyond its involvement in the form factor expansion (\ref{resness}), of the ``expectation value'' $\bra\sigma_\pm\ket_{\rho_{\rm ness}}$ would not be entirely clear -- such expectation values are usually determined from the large-distance asymptotic under the condition of conformal normalization at small distances.

An important conclusion, however, is that in general, both for the order and disordered regimes, we expect to have oscillatory terms in $\log(mx)$ with frequencies that are multiple of $2\gamma$. It would be interesting to understand more fully the large-distance expansion.

\subsubsection{``Non-equilibrium KMS relation'' and analytic properties}

In an equilibrium Gibbs state at temperature $\beta^{-1}$, the two-point function of fermion fields $G_\beta(x,\tau)=\bra\psi(x,\tau)\psi(0,0)\ket_{\rho_\beta}$ at imaginary time $t=-i\tau,\,\tau\in\R$ satisfies the quasi-periodicity relation (KMS relation, see for instance \cite{Kapusta})
\[
	G_\beta(x,\tau) = -G_\beta(x,\tau+\beta).
\]
This relation is obtained by using the cyclic property of the trace and the relation $e^{\beta H}\psi(x,\tau)e^{-\beta H} = \psi(x,\tau+\beta)$. It is at the basis of the description of the Gibbs state in terms of Euclidean (imaginary-time) fields on a cylinder of circumference $\beta$ (for fermions: with anti-periodic conditions). This relation is however broken out of equilibrium. In the case of the present free fermion theory there is a straightforward ``nonequilibrium KMS relation''. Recall the mode expansion (\ref{psimodes}) for the fermion fields. We may define non-local fermion operators by integration over rapidity half-lines,
\beqa
	\psi^{l,r}(x,t) &=& \frc12 \sqrt{\frc m\pi} \int_{\theta\gtrless 0} d\theta\,e^{\theta/2}
	\lt(a(\theta)\,e^{ip_\theta x - iE_\theta t}
	+ a^\dag(\theta)\,e^{-ip_\theta x + iE_\theta }\rt) \n
	\b\psi^{l,r}(x,t) &=& -\frc i2 \sqrt{\frc m\pi} \int_{\theta\gtrless 0} d\theta\,e^{-\theta/2}
	\lt(a(\theta)\,e^{ip_\theta x - iE_\theta t}
	- a^\dag(\theta)\,e^{-ip_\theta x + iE_\theta t}\rt).
	\label{psiveemodes}
\eeqa
Then, with $G_{\rm ness}(x,\tau) = \bra\psi(x,\tau)\psi(0,0)\ket_{\rho_{\rm ness}}$ and $G_{\rm ness}^{l,r}(x,\tau) = \bra\psi^{l,r}(x,\tau)\psi(0,0)\ket_{\rho_{\rm ness}}$, the non-equilibrium KMS relation for two-point fermion correlation functions is $G_{\rm ness}(x,\tau) = - G_{\rm ness}^{l}(x,\tau+\beta_{l})-G_{\rm ness}^{r}(x,\tau+\beta_{r})$. This is a non-local relation, hence does not ({\em a priori}) have a natural geometric interpretation.

Yet when considering instead the two-point function
\[
	g_{\rm ness}(x,\tau) := \bra\psi(x,\tau)\mu_+(0,0)\ket_{\rho_{\rm ness}}
\]
such a generalized KMS relation is useful as it allows us to establish analytic properties of form factors. Indeed, similar arguments, using the exchange relations with the disorder field $\mu_+$ (recall that it has a branch cut towards its right), lead in the non-equilibrium steady state to the non-local relations
\beq\label{nessKMS}
	g_{\rm ness}(x,\tau) = {\rm sgn}(x)\lt(
	g_{\rm ness}^l(x,\tau+\beta_l)+
	g_{\rm ness}^r(x,\tau+\beta_r)\rt).
\eeq
Here we can use a form factor expansion like (\ref{ffexp}), since the field $\psi$ one has only one-particle form factors that are nonzero. In particular, in the present case of the non-equilibrium steady state (\ref{Wness}), integrals exist on the real line, and the large-distance expansion can be obtained exactly. The non-equilibrium KMS relation (\ref{nessKMS}) combined with the form factor expansion then leads to nontrivial conditions on the analyticity properties of form factors; in particular, it implies the presence of a branch cut, with prescribed jump. As we show in Subsection \ref{ssectKMS}, this is in agreement with (\ref{jump}). 

\section{Calculations}

\subsection{Mixing} \label{ssectproof}

First, note that the pure-state limit of form factors can be defined by taking the limit $V(\theta)\to\infty$ (uniformly on $\theta$) of mixed-state form factors;  recall (\ref{WV}). By using the cyclic property of the trace in order to bring all operators $a(\theta)$ to the left while keeping all operators $a^\dag(\theta)$ to the right of $\Or$, we see that
\beq\label{pure}
	\lim_{V\to\infty} f^{\rho;\Or}_{+,\ldots,+,-,\ldots,-}(\theta_1,\ldots,\theta_j,\theta_{j+1},\ldots,\theta_N) =
	\bra \theta_N,\ldots,\theta_{j+1}|\Or|\theta_1,\ldots,\theta_j\ket
\eeq
where there are $j$ plus signs and $N-j$ minus signs as indices on the left-hand side. This holds for all values of rapidities such that $\theta_j\neq\theta_k$ for $j\neq k$, and thanks to (\ref{exch}) we may evaluate the limit for other choices of signs. This suggests that matrix elements of $\Or$ in ${\cal H}$ with rapidities both on the right and on the left be gathered into the function
\beq\label{ffH}
	f^{\Or}_{\ep_1,\ldots,\ep_N}(\theta_1,\ldots,\theta_N):=
	\lim_{V\to\infty} f^{\rho;\Or}_{\ep_1,\ldots,\ep_N}(\theta_1,\ldots,\theta_N).
\eeq
In particular, the usual form factors, with excited states on the right and the vacuum on the left, are $f^\Or(\theta_1,\ldots,\theta_N) = f^{\Or}_{+,\ldots,+}(\theta_1,\ldots,\theta_N)$.

It will be convenient in order to express formally our result to use a notation for form factors with general Liouville states $|A\ket^\rho$:
\beq\label{ffA}
	f^{\rho;\Or}\big[|A\ket^\rho\big] = {}^\rho
	\bra\vac|\Or^\ell|A\ket^\rho.
\eeq
We will also use the notation
\beq\label{ffAzero}
	f^{\Or}\big[|A\ket^\rho\big] := \lim_{V\to\infty} f^{\rho;\Or} [|A\ket^\rho\big].
\eeq

\begin{propo}\label{propoU}
There is a linear map
\beqa
	\frak{U}\ :\ \End({\cal H}) &\to& \End({\cal H}) \n
	\Or &\mapsto& \frak{U}(\Or) \label{tmap}
\eeqa
satisfying
\beq\label{Udag}
	\frak{U}(\Or^\dag) = \frak{U}(\Or)^\dag,
\eeq
and an operator $\lU$ on $\liou_\rho$, such that for all $|A\ket^\rho\in\liou_\rho$ and all $\Or\in\End({\cal H})$,
\beq\label{mapU}
	|\frak{U}(\Or)\ket^\rho = {\lU}|\Or\ket^\rho,
\eeq
and
\beq\label{t-map1}
	f^{\rho;\Or}\big[|A\ket^\rho\big] = f^{\frak{U}(\Or)}\big[|A\ket^\rho\big] = f^{\Or}\big[\lU^\dag |A\ket^\rho\big].
\eeq
We have
\beq\label{lU}
	{\lU} = \exp\lt[\int d\theta\,\frc{{\la}_-(\theta){\la}_+(\theta)}{1+e^{W(\theta)}}\rt].
\eeq
Further, if $\Or$ is a local fermionic descendant of the identity (i.e.~the identity itself, the fermion fields $\psi(x)$ and $\b\psi(x)$, their $x$-derivatives, or normal-ordered products of such), then $\frak{U}(\Or)$ is a linear combination of such local fermionic descendants with equal and lower numbers of fermion fields.
\end{propo}

In order to prove Proposition \ref{propoU}, let us consider the operators
\beq
	\Or = a^\dag(\theta_n)\cdots a^\dag(\theta_{k+1}) a(\theta_{k})
	\cdots a(\theta_1) =:  \prod_i a^{\ep_i}(\theta_i)
\eeq
for fixed $k,n$. These, for all nonnegative integers $n\geq k$, form a basis in ${\rm End}({\cal H})$ according to our assumptions.

Recall the canonical creation and annihilation operators $\la_\ep(\theta)$ and $\la_\ep^\dag(\theta)$ on $\liou_\rho$, with anti-commutation relations (\ref{aliou}). Let us introduce the normal-ordering operation $\nl\cdot \nl$ on $\liou_\rho$ which takes the operators $\la_\ep(\theta)$ to the right of the operators $\la_\ep^\dag(\theta)$ in the usual way. Then, using (\ref{aa}), we have
\beqa\label{ovac}
	\nl \Or^\ell \nl|\vac\ket^\rho &=& \prod_{i}
	\frc{\la_{\ep_i}^\dag(\theta_i)}{
	C_{-\ep_i}(\theta) }|\vac\ket^\rho\\
	{}^\rho\bra\vac|\nl \Or^\ell \nl &=& {}^\rho\bra\vac|\prod_{i}
	\frc{\la_{-\ep_i}(\theta_i)}{
	C_{\ep_i}(\theta)}
	\label{vaco}
\eeqa
where
\beq\label{Cep}
	C_\ep(\theta):= 1+e^{\ep W(\theta)}.
\eeq
Using (\ref{aliou}), (\ref{aliou2}) and (\ref{vaco}) we immediately see that
\beq
	{}^\rho\bra\vac|\nl \Or^\ell\nl|\theta_1,\ldots,\theta_j,\theta_{j+1},\ldots,
	\theta_N\ket_{+,\ldots,+,-,\ldots,-}^\rho =
	\bra\theta_N ,\ldots,\theta_{j+1}|\Or|\theta_1,\ldots,\theta_j\ket
\eeq
(in particular, both sides are nonzero if and only if $k=j$ and $n=N$). This implies (see (\ref{ffAzero}))
\beq\label{eq1}
	f^{\Or}\big[|A\ket^\rho\big] = {}^\rho\bra\vac|\nl \Or^\ell\nl|A\ket^\rho
\eeq
for all $A\in {\rm End}({\cal H})$.

From (\ref{lU}), we find
\beq\label{Uconj}
	\lU \la_\ep^\dag(\theta) \lU^{-1} = \la_\ep^\dag(\theta) +
	\ep \frc{C_{-\ep}(\theta)}{C_{+}(\theta)} \la_{-\ep}(\theta).
\eeq
Hence, using (\ref{ovac}),
\beqa
	\lU \nl \Or^\ell\nl |\vac\ket^\rho &=&
	\prod_{i=n}^{k+1}\lt(
	\frc{\la_{+}^\dag(\theta_i)}{
	C_{-}(\theta_i) } +
	\frc{\la_{-}(\theta_i)}{C_{+}(\theta_i)}\rt)
	\prod_{i=k}^1\lt(
	\frc{\la_{-}^\dag(\theta_i)}{
	C_{+}(\theta_i) } 
	- \frc{\la_{+}(\theta_i)}{C_{+}(\theta_i)}\rt)|\vac\ket^\rho\n
	&=& \prod_{i=n}^{k+1}\lt(
	\frc{\la_{+}^\dag(\theta_i)}{
	C_{-}(\theta_i) } +
	\frc{\la_{-}(\theta_i)}{C_{+}(\theta_i)}\rt)
	\prod_{i=k}^1\lt(
	\frc{\la_{-}^\dag(\theta_i)}{
	C_{+}(\theta_i) } 
	+ \frc{\la_{+}(\theta_i)}{C_{-}(\theta_i)}\rt)|\vac\ket^\rho\n
	&=& \Or^\ell |\vac\ket^\rho = |\Or\ket^\rho\label{eq1.5}
\eeqa
where we used the facts that $\{\la_-^\dag(\theta),\la_+(\theta')\}=0$ and that $\la_+(\theta)|\vac\ket^\rho=0$. Taking (\ref{mapU}) as a definition of the map $\frak{U}$, we have found
\beq\label{eq2}
	\Or^\ell|\vac\ket^\rho = \nl \frak{U}(\Or)^\ell\nl|\vac\ket^\rho.
\eeq

From (\ref{otherside}) we observe that $\lt({}^\rho \bra A|\frak{U}(\Or)\ket^\rho\rt)^* = {}^\rho \bra \t A|\frak{U}(\Or)^\dag\ket^\rho$ where $\t A$ is obtained by expressing $|A\ket^\rho$ in the Liouville particle basis, and in every term conjugating the coefficient, inverting the order of the rapidities, and flipping the particle types (note that this does not give $|A^\dag\ket^{\rho}$). Since $\lU$ is invariant under these operations, we find $\lt({}^\rho \bra A|\frak{U}(\Or)\ket^\rho\rt)^* = \lt({}^\rho \bra A|\lU|\Or\ket^\rho\rt)^*=  {}^\rho \bra \t A|\lU|\Or^\dag\ket^\rho = {}^\rho \bra \t A|\frak{U}(\Or^\dag)\ket^\rho$, which implies (\ref{Udag}). In particular, from (\ref{eq2}), we then have
\beq\label{eq3}
	{}^\rho\bra\vac|\Or^\ell = {}^\rho\bra\vac|\nl\frak{U}(\Or)^\ell\nl
\eeq

Putting together (\ref{eq1}) and (\ref{eq3}) we obtain
\beq
	f^{\frak{U}(\Or)}\big[|A\ket^\rho\big] =
	 {}^\rho\bra\vac|\nl \frak{U}(\Or)^\ell\nl|A\ket^\rho =
	 {}^\rho\bra\vac|\Or^\ell|A\ket^\rho
\eeq
which shows the first equation in (\ref{t-map1}). The second is obtained similarly but using the hermitian conjugate of (\ref{eq1.5}),
\beq
	f^{\Or}\big[\lU^\dag|A\ket^\rho\big] =
	 {}^\rho\bra\vac|\nl \Or^\ell\nl \lU^\dag |A\ket^\rho =
	 {}^\rho\bra\vac|\Or^\ell|A\ket^\rho.
\eeq

Finally, in order to show the last part of the proposition, we only need to show locality (it is clear that $\lU$ reduces the number of fermion operators, and in fact that it preserves the parity of this number). For this, we note that using (\ref{aa}) and (\ref{Uconj}), given any {\em linear} combination $A = \int d\theta\,\lt( u(\theta) a^\dag(\theta) + v(\theta) a(\theta)\rt)$, we find
\beq
	\lU^{-1}A^\ell \lU = A^\ell + {\bf R},\quad
	\lU^{-1}A^r \lU = A^r + {\bf R}(-1)^{n^\ell - n^r}
\eeq
for some operator ${\bf R}$ (that depends on $u$ and $v$). Hence,
\beq
	\lU^{-1} (A^\ell -  A^r(-1)^{n^\ell-n^r}) \lU = A^\ell-
	A^r(-1)^{n^\ell-n^r}.
\eeq
That is, with $[A,\cdot]$ being a commutator if $\cdot$ is bosonic and an anti-commutator if it is fermionic, we deduce
\beq
	|[A,\frak{U}(\Or)]\ket^\rho = (A^\ell - A^r (-1)^{n^\ell-n^r})
	|\frak{U}(\Or)\ket^\rho = \lU (A^\ell - A^r (-1)^{n^\ell-n^r})
	|\Or\ket^\rho = |\frak{U}([A,\Or])\ket^\rho.
\eeq
This gives (\ref{cpor}). Let us now take $\Or = \Or(x')$ to be a normal-ordered product of fermion fields at the point $x'$. Then $[\psi(x),\Or(x')]=[\psi^\dag(x),\Or(x')]=0$  for all $x\neq x'$. Hence also $[\psi(x),\frak{U}(\Or(x'))]=[\psi^\dag(x),\frak{U}(\Or(x'))]=0$  for all $x\neq x'$. Given that $\frak{U}(\Or(x'))$ is composed of normal-ordered terms each with finitely many factors of creation / annihilation operators, it must be a linear combination of normal-ordered product of fermion fields at the point $x'$, plus possibly a term proportional to the identity. By translation covariance, this latter term is independent of $x'$. It is clear from the form of $\lU$ that the term with the highest number of fermion field is the original field itself. By dimensional analysis, we have
\begin{eqnarray}\label{def44}
\frak{U}(\Or(x)) = \Or(x) + \sum_{\Psi:d_\Psi< d_\Or} m^{d_\Or-d_\Psi}F^{\Psi}_{\Or} \Psi(x)
\end{eqnarray}
where $\Psi$ are normal-ordered products of fermion fields or the identity field (not involving explicitly the mass), $d_\Psi$ are their dimensions and $d_\Or$ is the dimension of $\Or$; the sum is finite. In the limit where $V\to\infty$, the coefficients $F_\Or^\Psi$ tend to zero.

\subsection{Mixed-state form factors of twist fields} \label{ssecttwist}

We prove Proposition \ref{conj} by establishing a system of functional differential equations for the one- and two-particle form factors of the field $\mu_+$ and $\sigma_+$ (something similar can be done for $\mu_-,\, \sigma_-$), and by showing that (\ref{cff}), (\ref{ch}) satisfy it. The differential equations are obtained using, in particular, the fact that Wick's theorem, as described after Proposition \ref{conj}, applies for mixed-state form factors with higher numbers of particles (and also for traces with insertion of one twist field and many creation / annihilation operators). Wick's theorem is a consequence of the fact that the twist fields $\sigma$ and $\mu$ are simply related to normal-ordered exponentials of bilinear expressions in the creation / annihilation operators. For instance the order field $\sigma_+$ is $\bra\sigma\ket$ times
\beq\label{genform}
	:\exp\lt[\sum_{\ep_1,\ep_2}\int d\theta_1 d\theta_2\, F_{\ep_1,\ep_2}(\theta_1,\theta_2) a^{\ep_1}(\theta_1)a^{\ep_2}(\theta_2)
	\rt]:
\eeq
where $F_{\ep_1,\ep_2}(\theta_1,\theta_2) = -\frc1{2 \bra\sigma\ket} f^{\sigma_+}_{\ep_1,\ep_2}(\theta_1,\theta_2) $ are essentially the matrix elements on the Hilbert space. Hence the equations we derive hold in general for operators of the form (\ref{genform}). The differential equations are nonlinear first-order differential equations for the form factors seen as functionals of the function $W:\theta\mapsto W(\theta)$ that characterizes the density matrix. With the initial condition at $V(\theta)={\rm Re}(W(\theta))=\infty$, given by the matrix elements of the twist fields on the Hilbert space, the solution is unique. Indeed, these matrix elements fully characterize the kernel $F_{\ep_1,\ep_2}(\theta_1,\theta_2)$ in the bilinear expression.

\subsubsection{Nonlinear functional differential system of equations}

Let (see (\ref{Q}))
\beq\label{tf}
	\t f_\ep(\theta) := Q_{\ep}(\theta)\frc{\Tr \lt(\rho
	\,\mu_+\,a^\ep(\theta)\rt)}{
	\Tr \lt(\rho
	\,\sigma_+\rt)} ,\quad
	\t f_{\ep_1,\ep_2}(\theta_1,\theta_2) :=
	Q_{\ep_1,\ep_2}(\theta_1,\theta_2)
	\frc{\Tr \lt(\rho
	\,\sigma_+\,a^{\ep_1}(\theta_1)a^{\ep_2}(\theta_2)\rt)}{
	\Tr \lt(\rho
	\,\sigma_+\rt)}
\eeq
and similarly for higher numbers of insertions of creation and annihilation operators. Using $\delta \rho / \delta W(\beta) = -a^\dag(\beta)a(\beta) \,\rho$, we find
\[
	\frc{\delta}{\delta W(\beta)}
	\frc{\Tr \lt(\rho
	\,\mu_+\,a^\ep(\theta)\rt)}{
	\Tr \lt(\rho
	\,\sigma_+\rt)} = - \frc{\Tr \lt(\rho
	\,\mu_+\,a^\ep(\theta) a^\dag(\beta)a(\beta)\rt)}{
	\Tr \lt(\rho
	\,\sigma_+\rt)} +
	\frc{\Tr \lt(\rho
	\,\mu_+\,a^\ep(\theta)\rt)}{
	\Tr \lt(\rho
	\,\sigma_+\rt)} \frc{\Tr \lt(\rho
	\,\sigma_+\,a^\dag(\beta)a(\beta)\rt)}{
	\Tr \lt(\rho
	\,\sigma_+\rt)}.
\]
Simplifying the right-hand side through Wick's theorem,
\[
	\t f_\ep(\theta) \t f_{+,-}(\beta,\beta) - \t f_{\ep,+,-}(\theta,\beta,\beta)=\t f_+(\beta) \t f_{\ep,-}(\theta,\beta) - \t f_-(\beta) \t f_{\ep,+}(\theta,\beta),
\]
this implies, from the definition (\ref{Q}),
\beq\label{sys1}
	\lt(\frc{\delta}{\delta W(\beta)} + \frc{\ep\delta(\beta-\theta)}{
	1+e^{\ep W(\beta)}}\rt) \t f_\ep(\theta) =
	\frc{
	\t f_+(\beta) \t f_{\ep,-}(\theta,\beta) -
	\t f_-(\beta) \t f_{\ep,+}(\theta,\beta)
	}{4\cosh^2\frc{W(\beta)}2}.
\eeq
Likewise, differentiating $\Tr \lt(\rho
	\,\sigma_+\,a^{\ep_1}(\theta_1)a^{\ep_2}(\theta_2)\rt)/
	\Tr \lt(\rho
	\,\sigma_+\rt)$, we find
\beqa
	\lefteqn{\lt(\frc{\delta}{\delta W(\beta)} + \frc{\ep_1\delta(\beta-\theta_1)}{
	1+e^{\ep_1 W(\beta)}}+
	\frc{\ep_2\delta(\beta-\theta_2)}{
	1+e^{\ep_2 W(\beta)}}\rt) \t f_{\ep_1,\ep_2}(\theta_1,\theta_2)}
	\qquad\qquad \n &= &
	\frc{
	\t f_{\ep_1,+}(\theta_1,\beta) \t f_{\ep_2,-}(\theta_2,\beta) -
	\t f_{\ep_1,-}(\theta_1,\beta) \t f_{\ep_2,+}(\theta_2,\beta)
	}{4\cosh^2\frc{W(\beta)}2}.
	\label{sys2}
\eeqa
Equation (\ref{sys2}) gives a continuous family of non-linear first-order functional differential equations for the $W$-functionals $\t f_{++}(\theta_1,\theta_2)$, $\t f_{+-}(\theta_1,\theta_2)$, $\t f_{-+}(\theta_1,\theta_2)$ and $\t f_{--}(\theta_1,\theta_2)$ ($\theta_1,\theta_2\in\R$). Once this system of equations is solved, the solution can be fed into (\ref{sys2}) to provide a continuous family of linear first-order differential equations for the $W$-functionals $\t f_+(\theta)$ and $\t f_-(\theta)$ for all $\theta\in \R$. The continuous families can be slightly simplified using anti-symmetry under interchange of the particles (which is consistent with (\ref{sys1}) and (\ref{sys2})).

This system of equations can be translated into a system for the mixed-state form factors themselves using (\ref{Off}), since we have, using the notation $f_\ep(\theta) := \bra\sigma\ket_\rho^{-1} f_\ep^{\rho;\mu_+}(\theta)$, $f_{\ep_1,\ep_2}(\theta_1,\theta_2) := \bra\sigma\ket_\rho^{-1} f_{\ep_1,\ep_2}^{\rho;\sigma_+}(\theta_1,\theta_2)$,
\beq\label{tff}
	\t f_\ep(\theta) =
	f_\ep(\theta),\quad
	\t f_{\ep_1,\ep_2}(\theta_1,\theta_2) =
	f_{\ep_1,\ep_2}(\theta_1,\theta_2) +
	\lt(1+e^{-\ep_1 W(\theta_1)}\rt)
	\delta_{\ep_1+\ep_2,0}\,\delta(\theta_1-\theta_2).
\eeq
This gives
\beqa\label{sys1p}
	\frc{\delta f_\ep(\theta)}{\delta W(\beta)}  &=&
	\frc{
	f_+(\beta) f_{\ep,-}(\theta,\beta) -
	f_-(\beta) f_{\ep,+}(\theta,\beta)
	}{4\cosh^2\frc{W(\beta)}2}\\
	\frc{\delta  f_{\ep_1,\ep_2}(\theta_1,\theta_2)}{\delta W(\beta)}
	&= &
	\frc{
	 f_{\ep_1,+}(\theta_1,\beta) f_{\ep_2,-}(\theta_2,\beta) -
	 f_{\ep_1,-}(\theta_1,\beta) f_{\ep_2,+}(\theta_2,\beta)
	}{4\cosh^2\frc{W(\beta)}2}.
	\label{sys2p}
\eeqa
We see that the delta-function terms have all disappeared. 

\subsubsection{Unicity}

From the trace definition of mixed-state form factors, or more explicitly from the expression (\ref{t-map1}) of mixed-state form factors in terms of ordinary form factors, we can deduce the form of the large-${\rm Re}(W(\theta))$ expansion. We have
\[
	f_{\{\ep\}}(\{\theta\}) =
	f_{\{\ep\}}^{(0)}(\{\theta\}) + \int d\beta_1 \,e^{-W(\beta_1)}
	\, f_{\{\ep\}}^{(1)}(\{\theta\},\beta_1) + \int d\beta_1 d\beta_2\,
	e^{-W(\beta_1)-W(\beta_2)}
	\, f_{\{\ep\}}^{(2)}(\{\theta\},\beta_1,\beta_2) + \ldots 
\]
where ${\{\ep\}} = \ep$ or $\ep_1,\ep_2$, and $\{\theta\} = \theta$ or $\theta_1,\theta_2$ $\cdot = \ep_1,\ep_2$. Then, equations (\ref{sys1p}), (\ref{sys2p}) provide uniquely the solutions order by order, once the zeroth order has been fixed. For instance,
\[
	f^{(1)}_{\ep_1,\ep_2}(\theta_1,\theta_2,\beta) =
	f_{\ep_1,+}^{(0)}(\theta_1,\beta) f_{\ep_2,-}^{(0)}(\theta_2,\beta) -
	 f_{\ep_1,-}^{(0)}(\theta_1,\beta) f_{\ep_2,+}^{(0)}(\theta_2,\beta).
\]
Since the expressions for mixed-state form factors given in Proposition \ref{conj} agree with the usual pure-state form factors at ${\rm Re}(W(\theta))=\infty$, then we only have to verify that they satisfy the system of equations (\ref{sys1p}), (\ref{sys2p}) in order to prove that they are correct.

\subsubsection{Solution}

In order to verify that the expressions for mixed states form factors of Proposition \ref{conj} give rise to a solution to (\ref{sys1p}) and (\ref{sys2p}), it is sufficient to take $\theta_1$ and $\theta_2$ in their analyticity region, as the system of equations (\ref{sys1p}), (\ref{sys2p}) is itself analytic. Hence we consider $\theta_1\in I^+_{\ep_1}$ and $\theta_2\in I^+_{\ep_2}$, see (\ref{strip}). We then have
\beqa
	f_\ep(\theta) &=& \frc1{\sqrt{i}} h_\ep^+(\theta) \n
	f_{\ep_1,\ep_2}(\theta_1,\theta_2) &=&
	h_{\ep_1}^+(\theta_1)h_{\ep_2}^+(\theta_2)\,
	\lt( \tanh\frc{\theta_2-\theta_1}2\rt)^{\ep_1\ep_2} \n
	\frc{\delta}{\delta W(\beta)} h_\ep^+(\theta)
	&=& \frc{\ep}{2\pi i} \frc{1}{\sinh W(\beta)}
	\frc1{\sinh(\theta-\beta)} h_\ep^+(\theta).
\eeqa

Consider (\ref{sys1p}). On the left-hand side, we find
\[
	\frc{\ep}{2\pi i} \frc1{\sinh W(\beta)} \frc1{\sinh(\theta-\beta)}
	\frc1{\sqrt{i}}h_\ep^+(\theta)
\]
and on-the right-hand side, using (\ref{hphm}),
\[
	-\frc1{4\pi i} \frc1{\sinh W(\beta)} \lt(
	\lt(\tanh\frc{\beta-\theta}2\rt)^{-\ep} -
	\lt(\tanh\frc{\beta-\theta}2\rt)^{\ep}\rt)
	\frc1{\sqrt{i}} h_\ep^+(\theta).
\]
These are equal thanks to the relation
\[
	\frc1{\sinh x} = \frc12\lt(\coth \frc x2 - \tanh \frc x2\rt).
\]
Similarly, consider (\ref{sys2p}). On the left-hand side, we find
\[
	\frc1{2\pi i} \frc1{\sinh W(\beta)}\lt(
	\frc{\ep_1}{\sinh(\theta_1-\beta)} + \frc{\ep_2}{\sinh(\theta_2-\beta)}
	\rt) \lt(
	\tanh\frc{\theta_2-\theta_1}2\rt)^{\ep_1\ep_2}
	h_{\ep_1}(\theta_1) h_{\ep_2}(\theta_2) 
\]
whereas on the right-hand side, again using (\ref{hphm}),
\beqa
	&& -\frc1{4\pi i} \frc1{\sinh W(\beta)}\; \times \n &&  \times\; \lt(
	\lt(\tanh\frc{\beta-\theta_1}2\rt)^{\ep_1}
	\lt(\tanh\frc{\beta-\theta_2}2\rt)^{-\ep_2}
	-
	\lt(\tanh\frc{\beta-\theta_1}2\rt)^{-\ep_1}
	\lt(\tanh\frc{\beta-\theta_2}2\rt)^{\ep_2}
	\rt) \;\times \n && \times \;h_{\ep_1}(\theta_1) h_{\ep_2}(\theta_2). 
\eeqa
Again, these are equal thanks to the relations
\beqa
	\lefteqn{
	\lt(\frc1{\sinh(\theta_1-\beta)}+\frc1{\sinh(\theta_2-\beta) }\rt)
	\tanh\frc{\theta_2-\theta_1}2} \qquad\qquad && \n
	&=& \frc12\lt(\tanh\frc{\beta-\theta_2}2\coth\frc{\beta-\theta_1}2
	- \tanh\frc{\beta-\theta_1}2\coth\frc{\beta-\theta_2}2\rt) \n
	\lefteqn{
	\lt(\frc1{\sinh(\theta_1-\beta)}-\frc1{\sinh(\theta_2-\beta) }\rt)
	\coth\frc{\theta_2-\theta_1}2} \qquad\qquad&& \n
	&=& \frc12\lt(\coth\frc{\beta-\theta_2}2\coth\frc{\beta-\theta_1}2
	- \tanh\frc{\beta-\theta_1}2\tanh\frc{\beta-\theta_2}2\rt) \no
\eeqa
The latter relations can be verified by looking at both sides as analytic functions of $\beta$. In the first relation, both sides have poles at $\beta=\theta_1$ and $\beta=\theta_2$ with residues $\tanh\frc{\theta_1-\theta_2}2$, and in the second the residues are $\pm\coth\frc{\theta_1-\theta_2}2$  respectively. In all cases, both sides change sign under $\beta\mapsto\beta+i\pi$ and no other poles than those mentioned are found in any strip of width $i\pi$.

\subsection{A thermodynamic argument} \label{ssectthermo}

Consider the one-particle mixed-state form factor
\beq\label{wer}
	f_{+}^{\rho;\mu_+}(\theta) = 
	(1+e^{-W(\theta)})\frc{\Tr\lt(\rho \,\mu_+ a^\dag(\theta)\rt)}{
	\Tr\lt(\rho\rt)}.
\eeq
The idea for the thermodynamic argument is to evaluate the traces by taking the product over the one-particle subspaces of all rapidities. This is clear for the denominator since $\rho$ is diagonal:
\[
	\frc{(1+e^{-W(\theta)})}{\Tr(\rho)} = \frc1{
	\prod_{\theta'\neq\theta} (1+e^{-W(\theta')})}.
\]

For the numerator, the argument is less clear, as $\mu_+$ is not diagonal. Nevertheless, we may imagine at each rapidity that we look at the region far to the right, where there is a cut. Looking at this region will be corresponding to a slight imaginary shift in the rapidity, to be implemented below. Looking at the other region should give a factor of 1 by the argument presented here.

Taking the product over the rapidities and taking into account the sign change due to the cut, this should give $\prod_{\theta'} (1-e^{-W(\theta')})$ on the numerator. However, the presence of the particle at the rapidity $\theta$ will affect the density of rapidities $\theta'$ over which we take the product, as well as what we obtain at $\theta'=\theta$. Considering a region at a distance $L$ to the right of the origin, we expect there to be a factor representing the effect of the branch point, given by
\[
	\bra\vac|\sigma_+|\theta,\theta'\ket e^{-ip_{\theta'} L}
	\propto \tanh\frc{\theta-\theta'}2 e^{-ip_{\theta'} L},
\]
in evaluating the trace on the $\theta'$ subspace, when one particle at $\theta'$ is present. The dominant contribution will be obtained when this factor is equal to 1; then in particular we do get $1-e^{-W(\theta')}$ in evaluating the trace. At $\theta'=\theta$ this is not possible as the tanh function is zero -- there the denominator is simply equal to 1. Hence we must evaluate
\[
	 \prod_{\theta'\neq \theta} \frc{1-e^{-W(\theta')}}{1+e^{-W(\theta')}}
	 = \exp\lt[ \sum_{\theta'\neq\theta} \log\lt(\tanh\frc{W(\theta')}2\rt)\rt]
\]
with a density obtained from
\[
	\frc1L = -\frc1{2\pi i L} d
	\log\lt(\tanh\frc{\theta-\theta'}2\rt)
	+ \frc1{2\pi} d p_{\theta'}.
\]
In fact, near to $\theta$ this argument fails, and the fact that we were looking at the far right region must be taken into account. This appears to mean that we must replace $\theta'$ by $\theta'-i{\bf 0}$ (thus making $e^{-ip_{\theta'} L}$ decaying at large $L$), and that we must the sum over all values of $\theta'$, including the value $\theta$.

Then, the part corresponding to $\Delta p_{\theta'}$ gives an infinite contribution. But if the twist field is at position $x$ instead of $0$, then we must replace $L$ by $L-x$, and this corresponds to the finite change
\[
	\exp\lt[x \int \frc{dp_{\theta'}}{2\pi} \log\lt(\coth\frc{W(\theta')}2\rt)\rt]
\]
in agreement with (\ref{transcov}), (\ref{E}).

The other part gives
\[
	\exp\lt[\frc1{2\pi i}\,\int d\theta' \frc{\p}{\p \theta'}
	\log\lt(\coth\frc{\theta-\theta'+i{\bf 0}}2\rt)\;
	\log\lt(\tanh\frc{W(\theta')}2\rt)\rt].
\]
This is in agreement with (\ref{ch}).

Of course, this argument is far from a satisfying derivation, but it does explain the main features of (\ref{ch}).

\subsection{Analytic structure from non-equilibrium KMS relation} \label{ssectKMS}

We re-write the equality (\ref{nessKMS}) using a form factor expansion like (\ref{ffexp}) (and simplifying common factors):
\[
	\sum_\ep \int d\theta\, e^{\theta/2}
	\frc{e^{i\ep p_\theta x -\ep E_\theta \tau}}{
	1+e^{-\ep W_{\rm ness}(\theta)}}
	h_{-\ep}^+(\theta)
	= {\rm sgn}(x)
	\sum_\ep \int d\theta\, e^{\theta/2}
	\frc{e^{i\ep p_\theta x -\ep \tau (E_\theta +W_{\rm ness}(\theta))}}{
	1+e^{-\ep W_{\rm ness}(\theta)}}
	h_{-\ep}^+(\theta).
\]
In order to establish the equality, we shift the $\theta$-contour by $\ep i\pi/2$ for $x>0$ and by $-\ep i\pi/2$ for $x<0$, taking care of surrounding the segment from 0 to $\pm\ep i\pi/2$. Let us denote by
\[
	g^r_{\ep}(\theta) := \frc{h_{-\ep}^+(\theta)}{1+e^{-\ep\beta_r E_\theta}},\quad
	g^l_{\ep}(\theta) := \frc{h_{-\ep}^+(\theta)}{1+e^{-\ep\beta_l E_\theta}}.
\]
Let us consider for instance the case $x>0$. Equality will be obtained if the $\ep=+$ and $\ep=-$ shifted contours cancel each other except at the points where the extra factor $e^{-\ep \tau W_{\rm ness}(\theta)}$ is ${\rm sgn}(x)=1$. That is, we require
\beq\label{cancel}
	e^{i\pi/4} g^{r,l}_+\lt(\theta+\frc{i\pi}2\rt)
	+e^{-i\pi/4} g^{r,l}_-\lt(\theta-\frc{i\pi}2\rt) = 0
\eeq
except for the values of $\theta$ satisfying
\[
	e^{i m\beta_{r,l} \sinh\theta} = 1
\]
where the functions $g^{r,l}_+\lt(\theta+\frc{i\pi}2\rt)$ and $g^{r,l}_-\lt(\theta-\frc{i\pi}2\rt)$ may have poles. One can check that with (\ref{ch}), these do have poles at the those values and that (\ref{cancel}) holds elsewhere. Further, equality requires that the parts of the contours surrounding the segments emanating from the origin also cancel. On the right-hand side, these four terms are
\beqa
	&& \int_{i\pi/2}^0 d\theta \,e^{\theta/2} g^r_+(\theta-{\bf 0})
	e^{ip_\theta x - E_\theta (\tau+\beta_r)} +
	\int_{0}^{i\pi/2} d\theta \,e^{\theta/2} g^l_+(\theta+{\bf 0})
	e^{ip_\theta x - E_\theta (\tau+\beta_l)} + \n
	&& +
	\int_{-i\pi/2}^0 d\theta \,e^{\theta/2} g^r_-(\theta-{\bf 0})
	e^{ip_\theta x - E_\theta (\tau+\beta_r)} +
	\int_{0}^{-i\pi/2} d\theta \,e^{\theta/2} g^l_-(\theta+{\bf 0})
	e^{ip_\theta x - E_\theta (\tau+\beta_l)}
\eeqa
and on the left-hand side they are the same except for setting $\beta_{r,l}$ to zero. Equality on the segment from $0$ to $i\pi/2$, and equality on the segment from $0$ to $-i\pi/2$, then give the two equations (for $\ep=\pm$ respectively)
\[
	g^r_\ep(\theta-{\bf 0}) e^{-\ep\beta_r E_\theta} -
	g^l_\ep(\theta+{\bf 0}) e^{-\ep\beta_l E_\theta} =
	g^r_\ep(\theta-{\bf 0})  -
	g^l_\ep(\theta+{\bf 0})
\]
which imply
\[
	\frc{g^r_\ep(\theta-{\bf 0})}{g^l_\ep(\theta-{\bf 0})}
	= \frc{1-e^{-\ep\beta_l E_\theta}}{1-e^{-\ep\beta_r E_\theta}}.
\]
This gives exactly (\ref{jump}) in the case of $h^+_-$ for $\ep=+$, and in the case of $h^+_+$ for $\ep=-$ (and with $W_{\rm ness}$).

A similar analysis can be done with $x<0$, giving a pole structure in agreement with (\ref{ch}) and showing that $h_+^+(\theta)$ has no jump through the segment from $0$ to $i\pi/2$, and that $h_-^+(\theta)$ has no jump through the segment from $0$ to $-i\pi/2$, again in agreement with our results.

\subsection{General solution as integral-operator kernel}\label{ssectint}

The nonlinear functional differential equation (\ref{sys2p}) presented in Subsection \ref{ssecttwist} is valid for general normal-ordered exponential fields of the form (\ref{genform}). However, the solution we presented is a special solution, for the twist fields $\sigma_+$. It is not clear {\em a priori} that a general solution will have the leg-factor structure found. Nevertheless, we may express the general solution of (\ref{sys2p}) as an integral-operator kernel constructed out of the functions $F_{\ep_1,\ep_2}(\theta_1,\theta_2)$ involved in (\ref{genform}).

We may proceed as follows. Denote $:e^S:$ the operator (\ref{genform}), with
\[
	S = \sum_{\ep_1,\ep_2}\int d\theta_1 d\theta_2\, F_{\ep_1,\ep_2}(\theta_1,\theta_2) a^{\ep_1}(\theta_1)a^{\ep_2}(\theta_2),
\]
and consider its two-particle mixed-state form factors, for convenience here taking the two-particle state on the left, ${}_{\ep_1,\ep_2}^{\hspace{4mm} \rho}\bra\theta_1,\theta_2|(:e^S:)^\ell |\vac\ket^\rho$. We have
\beqa
	{}_{\ep_1,\ep_2}^{\hspace{4mm} \rho}\bra\theta_1,\theta_2|
	(:e^S:)^\ell |\vac\ket^\rho &=&
	{}_{\ep_1,\ep_2}^{\hspace{4mm} \rho}\bra\theta_1,\theta_2|
	\lU\nl  e^{S^\ell}\nl |\vac\ket^\rho \n
	&=&
	{}_{\ep_1,\ep_2}^{\hspace{4mm} \rho}\bra\theta_1,\theta_2|
	\lU e^{\t S} |\vac\ket^\rho\n
	&=&
	{}_{\ep_1,\ep_2}^{\hspace{4mm} \rho}\bra\theta_1,\theta_2|
	e^{\lU \t S\lU^{-1}} |\vac\ket^\rho\no.
\eeqa
In the first step we used (\ref{eq2}). In the second step we use the left-action expressions (\ref{aa}) and the Liouville-space normal ordering in order to simplify the action on the vacuum, defining
\beq
	\t S = \sum_{\ep_1,\ep_2}\int \frc{d\theta_1 d\theta_2}{C_{-\ep_1}(\theta_1)C_{-\ep_2}(\theta_2)}
	F_{\ep_1,\ep_2}(\theta_1,\theta_2)\, \la^\dag_{\ep_1}(\theta_1)\la^\dag_{\ep_2}(\theta_2)
\eeq
(see (\ref{Cep})). In the last step we used $\lU|\vac\ket^\rho=0$. Using (\ref{Uconj}) we then obtain
\beq\label{ffas}
	{}_{\ep_1,\ep_2}^{\hspace{4mm} \rho}\bra\theta_1,\theta_2|
	(:e^S:)^\ell |\vac\ket^\rho =
	{}_{\ep_1,\ep_2}^{\hspace{4mm} \rho}\bra\theta_1,\theta_2|
	e^{D} |\vac\ket^\rho
\eeq
where
\beq\label{Dmat}
	D = \sum_{\ep_1,\ep_2}\int d\theta_1 d\theta_2\,
	F_{\ep_1,\ep_2}(\theta_1,\theta_2)\,
	\lt(\frc{\la^\dag_{\ep_1}(\theta_1)}{C_{-\ep_1}(\theta_1)}
	+\frc{\ep_1\,\la_{-\ep_1}(\theta_1)}{C_+(\theta_1)}\rt)
	\lt(\frc{\la^\dag_{\ep_2}(\theta_2)}{C_{-\ep_2}(\theta_2)}
	+\frc{\ep_2\,\la_{-\ep_2}(\theta_2)}{C_+(\theta_2)}\rt).
\eeq
On the right-hand side of (\ref{ffas}) we now have a matrix element of a pure exponential, so that we may use standard Bogoliubov-transformation techniques.

One way to implement such techniques is as follows. Consider a general canonical anti-commutation algebra with basis $b_j$, $b_j^*$, taking discrete indices $j=1,2,\ldots$ for simplicity, with $\{b^*_j,b_k\} = \delta_{j,k}$, other anti-commutators being zero. Denote by $|0\ket$ the associated Fock vacuum. Denote by $V$ the column vector divided into two blocks, whose elements are
\[
	V = (b_1,b_2,\ldots;\;b_1^*,b_2^*,\ldots)^T,\quad
	V^* = (b_1^*,b_2^*,\ldots;\;b_1,b_2,\ldots).
\]
For general 2-block by 2-block matrix $J$, we are interested in evaluating the matrix
\beq\label{mael}
	\frc{\bra 0 | V V^* e^{V^* J V}|0\ket}{
	\bra 0 | e^{V^* J V}|0\ket}
	= W:= \mato{cc} W_{11} & W_{12} \\ 0 & 0 \matf
\eeq
where on the right-hand side we extracted the block structure. In particular, for the form factor problem at hand, we are looking for $W_{12}$. We evaluate (\ref{mael}) by first solving a more general problem, writing
\beq
	\bra 0 | e^{V^* J V}|0\ket\, W = \lim_{\beta\to\infty} \Tr\lt(V V^* e^{V^* J V} e^{-\beta N}
	\rt),\quad N = \frc12 V^* \sigma_z V
\eeq
where $\sigma_z$ acts on the block space. The trace may be evaluated as usual by using its cyclic property, moving $V$ along one cycle, and the commutation relations
\[
	\{V,V^*\} = {\bf 1},\quad
	[N,V] = -\sigma_z V,\quad
	[V^* J V,V] = MV,\quad M = \sigma_x J^T\sigma_x - J.
\]
The result is
\[
	\lt(1+e^{\beta \sigma_z} e^M\rt)
	\Tr\lt(V V^* e^{V^* J V} e^{-\beta N}\rt) =
	e^{\beta\sigma_z} e^M \Tr\lt(e^{V^* J V} e^{-\beta N}\rt).
\]
We take the large-$\beta$ limit, keeping only the divergent terms proportional to $e^{\beta}$ on both sides. For this, we use $e^{\beta \sigma_z} \to e^{\beta}P_1$ where $P_1$ is the projector onto the first block. Hence we find
\[
	P_1 e^{M} W = P_1 e^M
	\;\Rightarrow\;
	(e^M)_{11} W_{12} = (e^M)_{12}
	\;\Rightarrow\;
	W_{12} =\lt((e^M)_{11}\rt)^{-1} (e^M)_{12}
\]
which is the standard expression.

We now apply this to our form factor problem. Using (\ref{Dmat}) and canonically normalized operators
\[
	\lb_\ep^*(\theta) =
	\frc{\sqrt{C_+(\theta)}}{C_{-\ep}(\theta)} \la^\dag_\ep(\theta),\quad
	\lb_\ep(\theta) = \frc1{\sqrt{C_+(\theta)}}\la_\ep(\theta)
\]
we find
\beq\label{Dmatnorm}
	D = \sum_{\ep_1,\ep_2}\int \frc{d\theta_1 d\theta_2}{
	\sqrt{C_+(\theta_1)C_+(\theta_2)}}\,
	F_{\ep_1,\ep_2}(\theta_1,\theta_2)\,
	\lt(\lb^\dag_{\ep_1}(\theta_1)
	+\ep_1\,\lb_{-\ep_1}(\theta_1)\rt)
	\lt(\lb^\dag_{\ep_2}(\theta_2)
	+\ep_2\,\lb_{-\ep_2}(\theta_2)\rt).
\eeq
This gives rise to the matrix of integral operators $J$ by identifying $D = V^*JV$ and then to $M$. In the 4 by 4 form taking into account the blocks discussed above as well as the internal particle-type $\ep$ block structure, we obtain
\beq
	M = \mato{cccc} x & -y & -y & -x \\
	z & x^t & x^t & -z \\
	-z & -x^t & -x^t & z \\
	x & -y & -y & -x \matf
\eeq
where the integral operators $x,y,z$ have kernels
\beqa
	x(\theta_1,\theta_2)
	&=& \frc{F_{+-}(\theta_1,\theta_2) - F_{-+}(\theta_2,\theta_1)
	}{\sqrt{C_+(\theta_1)C_+(\theta_2)}} \n
	y(\theta_1,\theta_2) &=&
	\frc{F_{++}(\theta_1,\theta_2)
	}{\sqrt{C_+(\theta_1)C_+(\theta_2)}} \n
	z(\theta_1,\theta_2) &=& \frc{F_{--}(\theta_1,\theta_2)
	}{\sqrt{C_+(\theta_1)C_+(\theta_2)}}
	 \no
\eeqa
and $x^t(\theta_1,\theta_2) = x(\theta_2,\theta_1)$ is the transposed kernel. We note that $M$ is nilpotent, $M^2 = 0$, so that its exponential can be calculated trivially,
\[
	e^M = 1+M.
\]
By direct calculations one then finds
\[
	Z:=W_{12} = \mato{cc}
	-(1+q)^{-1}\, y\, (1+x^t)^{-1} & -q\,(1+q)^{-1} \\ - q^t\,(1+q^t)^{-1}
	& (1+q^t)^{-1}\, z\, (1+x)^{-1} \matf,\quad
	q = x + y\,(1+x^t)^{-1}\, z.
\]
This gives the form factors via the kernel of this operator,
\[
	{}_{\ep_1,\ep_2}^{\hspace{4mm} \rho}\bra\theta_1,\theta_2|
	(:e^S:)^\ell |\vac\ket^\rho =
	\sqrt{C_+(\theta_1)C_+(\theta_2)}
	\, Z_{\ep_1,\ep_2}(\theta_1,\theta_2)
\]
where $\ep_{1,2}=+$ is on the first row / column and $\ep_{1,2}=-$ is on the second.

\section{Conclusion and discussion}

\subsection{Work done}

Generalizing the thermal (or finite-temperature) form factors of \cite{Ben1,Ben3}, we have defined mixed-state form factors and evaluated them explicitly in the Ising model, and we have shown how to obtain a full large-distance expansion for mixed-state two-point correlation functions of order and disorder fields from these form factors. Our method, like that of \cite{Ben1,Ben3}, differs in an important way from other methods more widely used in the literature on massive integrable models: we do not ``explicitly'' perform the trace defining the two-point function with ordinary form factors and matrix elements (necessitating a delicate cancelation of  divergencies), but rather we define mixed-state form factors as simpler traces, evaluated essentially using the cyclic property, and show that these traces can be used as building blocks for correlation functions, paralleling the ordinary vacuum K\"allen-Lehmann expansion. Our basic idea essentially follows from the GNS construction of $C^*$-algebras.

We evaluated mixed-state form factors using a novel technique, showing how to derive for them a system of non-linear functional-differential equations. Knowing the vacuum form factors, these equations provide uniquely all mixed state form factors up to normalization. These techniques appear similar to techniques used in classical integrable models in order to obtain bilinear differential equations for tau-functions, which are however usually associated to correlation functions instead of form factors (see for instance \cite{Babelon}). But we do not know yet if there is a full technical equivalence. Our new technique is a departure from the standard techniques based on solving a Riemann-Hilbert problem for vacuum or thermal form factors. It is more powerful as it does not require any strong analyticity property for the eigenvalues of the density matrix.

We have indicated how to apply our form factor expansions to some physical situations of interest: quantum quenches and thermal-flow non-equilibrium steady states. In particular, we have found a particular oscillating behaviour in $\log(mx)$ in the latter situation, with a frequency determined by the temperatures of the asymptotic baths.

Our form factor expansion can be trivially re-expressed as the determinant of a matrix linear integral operator, generalizing for instance the ideas of \cite{Babelon2} (as was done in \cite{Ben1} for the thermal case). From this perspective, we note that techniques of \cite{Izergin} for obtaining such representations of thermal correlation functions in the XY model are likely to be applicable as well to the case of general diagonal density matrices.

\subsection{Further developments}\label{ssectfurther}

There are many directions that the current work has left to be explored.

First, a development of both applications to quantum quenches and to the non-equilibrium energy-flow steady state are clearly desirable. In particular, this will require a full extension of our formalism to more general functions $W(\theta)$, including with discontinuities and with regions of negativity.

Second, and related to the above, it would be most interesting to understand what generalizes the ``quantization on the circle'' viewpoint of thermal correlation functions, and the Matsubara frequencies. From our analysis of the quantum quench and non-equilibrium steady state applications, it appears that the generalization will be based on a study of the singularity structure of the filling factors associated to the density matrix (bosonic or fermionic, depending on the semi-locality structure of the field considered). That is, there should be a correspondence between the spectrum in an alternative quantization scheme, where mixed-state correlation functions are reproduced by vacuum expectation values, and the singularity structure of these filling factors. It would also be interesting to see how these principles, here understood in the context of the Ising model, can be applied to more general QFT in mixed states.

Third, by the symmetry arguments of \cite{Fonseca1} (or generalizing the arguments of \cite{Babelon2}), one can easily see that the mixed-state correlation functions we considered satisfy a system of bilinear partial differential equations which are essentially the Hirota form of the sinh-Gordon equation. As was discovered in \cite{Gamsa}, the thermal form factors of the Ising model can be seen as the initial scattering data for the associated inverse scattering problem. We expect the same to hold in the cases of more general mixed states. In particular, it is possible that the choices of density matrices be in  correspondence with the possible initial conditions for the sinh-Gordon equation.

Fourth, it would be most interesting to analyze the large-time expansion of correlation functions. The aforementioned sinh-Gordon equation method may be of use, as well as, for instance, the virial expansion method of \cite{Reyes06}.

Fifth, we note that although we have not evaluated the normalization of form factors (and in the non-equilibrium steady-state case, some subtleties arise), it is likely that our techniques of deriving non-linear functional differential equations can be used similarly for the normalization, along with a careful cancellation of divergencies, in order to evaluate it exactly.

Finally, not only our techniques can be immediately applied to any other free-fermion (or free-boson) model, like the sine-Gordon model at the free-fermion point, but also we expect similar techniques to provide an interesting new approach to the study of thermal and more generally mixed-state correlation functions in integrable models with non-trivial scattering matrix. For instance, it is likely that one can obtain in a very efficient way the low-temperature expansion of thermal correlation functions, and more generally there is a hope, using thermodynamic arguments combined with exact functional-differential equations, to obtain exact mixed-state form factors.

\vspace{0.6cm}

\noindent {\bf Acknowledgments}\\
BD would like to thank Denis Bernard, John Cardy, Oleg Lisovyy, Robert Konik, Sergei Lukyanov, Gabor Tak\'acs and Alexei Tsvelik for discussions and comments, as well as the Simons Center for Geometry and Physics for hospitality and support during the last stages of this work.

\appendix

\section{Simple properties in the Liouville space}\label{appLiou}

\subsection{Left- and right-actions}

The Liouville left-action is defined in (\ref{lefta}). Similarly, to every $B\in\End{\cal H}$ one can define a right-action $B^r$ by $B^r|A\ket^\rho = |AB\ket^\rho$. The right-action linear map $B\mapsto B^r$ is an algebra anti-homomorphism $(AB)^r = B^rA^r$.  Using Definition (\ref{ip}), one can see that on conjugate vectors we have
\beq\label{aconj}
	{}^\rho\bra A|B^\ell = {}^\rho\bra B^\dag A|,\quad
		{}^\rho\bra A|B^r = {}^\rho\bra A\,\rho B^\dag \rho^{-1} |.
\eeq
Clearly, left- and right-action Liouville operators commute with each other, $A^\ell B^r = B^r A^\ell$.

\subsection{Hermitian conjugations}

It is a simple matter to translate more generally the Hermitian conjugation of operators on ${\cal H}$ onto that of operators on $\liou_\rho$ (as we mentioned, we denote the Hermitian conjugation in both cases by ${}^{\;\dag}$). In particular, conjugating the equations (\ref{aconj}), we find, for every $B\in\End({\cal H})$,
\beq\label{hl}
	\big(B^\ell\big)^\dag = \big(B^\dag\big)^\ell,
	\quad \big(B^r\big)^\dag = \lt(\rho B^\dag \rho^{-1}\rt)^r.
\eeq
That is, the Hermitian conjugation commutes with the left-action map, but not with the right-action map. Specializing to operators $a^\ep(\theta)$ and using
\beq\label{rhoa}
	\rho \,a^\ep(\theta)\,\rho^{-1} = e^{-\ep W(\theta)} a^\ep(\theta)
\eeq
as well as linearity of the right-action map, we obtain
\beq\label{hr}
	\big(a^\ep(\theta)^r\big)^\dag = e^{\ep W(\theta)}
	 a^{-\ep}(\theta)^r.
\eeq

\subsection{Creation and annihilation operators}

We can of course express the left- and right-actions of the Hilbert space creation and annihilation operators $a^\ep(\theta)$ in terms of the operators $\la_\ep(\theta)$, $\la_\ep^\dag(\theta)$ (\ref{aliou}). Let $(-1)^{n}=(-1)^{\int d\theta\, a^\dag(\theta) a(\theta)}\in\End({\cal H})$ be the operator that gives the parity of the number of Hilbert-space particles. Then $(-1)^{n^\ell-n^r}\in\End(\liou_\rho)$ is the operator that gives the parity of the number of Liouville particles, which can also be written $(-1)^{{\bf n}} = (-1)^{\sum_\ep\int d\theta\,\la_\ep^\dag(\theta)\la_\ep(\theta)/(1+e^{-\ep W(\theta)})}$. From (\ref{basis}) and (\ref{Q}), it is clear that we should have $a^\ep(\theta)^\ell = \frc{\la^\dag_\ep(\theta)}{1+e^{-\ep W(\theta)}} + q_\ep \la_{-\ep}(\theta)$ and $a^\ep(\theta)^r(-1)^{n^\ell-n^r}=  \frc{\la^\dag_\ep(\theta)}{1+e^{-\ep W(\theta)}} + \t q_\ep \la_{-\ep}(\theta)$, where in both cases, the second term must be present in order to account for the fact that no delta-function contact term is obtained in permuting Liouville particles, but that such contact terms appear in commuting Hilbert-space annihilation and creation operators. The values of $q_\ep$ and $\t q_\ep$ can be fixed uniquely using the Hermitian structure of $\liou_\rho$, more precisely (\ref{hl}) and (\ref{hr}) (which imply in particular $(n^\ell)^\dag = n^\ell$, $(n^r)^\dag = n^r$). We find
\beqa
	a^\ep(\theta)^\ell &=& \frc{\la^\dag_\ep(\theta)}{1+e^{-\ep W(\theta)}} + \frc{\la_{-\ep}(\theta)}{1+e^{\ep W(\theta)}} \n
	a^\ep(\theta)^r(-1)^{n^\ell-n^r} &=&  \frc{1}{1+e^{-\ep W(\theta)}}\lt(\la^\dag_\ep(\theta) + \la_{-\ep}(\theta)\rt).
	\label{aa}
\eeqa

\subsection{Hamiltonian and momentum operators}

Finally, we note that our choice (\ref{formrho}) of density matrix $\rho$ guarantees that space and time translation invariance hold as well in $\liou_\rho$ (however, of course, we do not have in general Poincar\'e invariance). Indeed, we can define operators
\beqa
	\lH &:=& H^\ell - H^r = \sum_\ep \int d\theta\,\ep\,E_\theta \,\frc{\la_\ep^\dag(\theta) \la_\ep(\theta)}{1+e^{-\ep W(\theta)}}\n
	\lP &:=& P^\ell - P^r = \sum_\ep \int d\theta\,\ep\,p_\theta \,\frc{\la_\ep^\dag(\theta) \la_\ep(\theta)}{1+e^{-\ep W(\theta)}}
\eeqa
which commute with each other, are Hermitian, and are diagonalized on the basis $|\theta_1,\ldots,\theta_N\ket_{\ep_1,\ldots,\ep_N}^\rho$, with in particular $\lH|\vac\ket^\rho = \lP|\vac\ket^\rho = 0$. For any ${\bf A}\in\End(\liou_\rho)$, space and time translations are given by
\beq
	{\bf A}(x,t) = e^{i\lH t-i\lP x} {\bf A} e^{-i\lH t +i\lP x}.
\eeq
This is in agreement with the left- and right-action maps: with $A(x,t) = e^{iHt-iPx}A e^{-iHt +iPx}$ we have
\beq
	A(x,t)^\ell = A^\ell(x,t),\quad
	A(x,t)^r = A^r(x,t)
\eeq
where we used the homomorphism and anti-homomorphism properties of the left- and right-action maps, respectively, as well as the fact that left- and right-action Liouville operators commute with each other. It is also in agreement with the correspondence between Hilbert space operators and Liouville space vectors:
\beq
	|A(x,t)\ket^\rho = e^{i\lH t-i\lP x}|A\ket^\rho.
\eeq

\section{Reality of correlation function}\label{appreal}

The order and disorder two-point correlation functions should be real functions. This is a non-trivial fact that encompasses information both about the hermitian conjugation properties of the fields, and about their locality. Reality is not explicit in the form factor expansion (\ref{corrsimple}). We provide a simple check here that the one-particle term is indeed real. The verification to higher particles is more involved and we defer it to another paper.

In order to verify that the one-particle order of the form factor expansion is real, we note that
\[
	K_\ep(\theta)^* = e^{\ep W(\theta)} K_{-\ep}(\theta).
\]
Hence, after changing variable to $\ep_j\mapsto -\ep_j$, the imaginary part of the one-particle integral on right-hand side of (\ref{corrsimple}) is proportional to
\beq\label{intrewq}
\sum_{\ep}
\int d\theta \,\lt(1-e^{-\ep W(\theta)}\rt)K_{\ep}(\theta)
e^{i\ep p_{\theta} |x| - i\ep E_{\theta} t}.
\eeq
The function $J_\ep(\theta) = (1-e^{-\ep W(\theta)}) K_\ep(\theta) = i^{\ep-1} g(\theta+i\ep {\bf 0})^{-2\ep}/(2\pi)$ can be analytically continued, and satisfies
\[
	J_+(\theta+i\pi/2) = -J_-(\theta-i\pi/2).
\]
Hence, shifting the $\theta$-contour by $i\ep \pi/2$ (no pole is crossed), we find that (\ref{intrewq}) is zero.

\section{Leading large-distance behavior of GGE correlation functions}
\label{appleading}

We concentrate on the GGE described by \eqref{WGGE}, obtained after a quench of the magnetic field \cite{Fagotti,Fagotti1,Fagotti2}.

Let us first consider the leading large-distance behavior in the disordered regime. The spin-spin correlation function $\t G(x,0)$ has an exponentially decaying factor  $e^{-x{\cal E}}$ controlled by ${\cal E}$ \eqref{E}. Other leading factors come from the one-particle contribution,
\[
\sum_{\ep} \frc 1 2 \int d\theta \,e^{i\ep p_{\theta} x}(1+\ep U(\theta))f^{\rho^{\sharp};\mu_+}_{\ep}(\theta) f^{\rho^{\sharp};\mu_-}_{-\ep}(\theta)
\]
where $U$ is given by  \eqref{UGGE}. There are poles at $\sinh\theta=\pm i \sqrt {m_0/m}$ and branch points at $\sinh\theta=\pm i  m_0/m$.  If $m_0>m$, then $\sqrt{m_0/m}<m_0/m$, hence the poles determine the least-decaying behavior. In this case, we deform the $\theta$ contours away from the real line, in the direction ${\rm sign}({\rm Im}(\theta))=\ep$ in which the form factors are analytic, all the way to ${\rm Im}(\theta) = \ep i\pi/2$. There, poles are found at $\theta^\star = \ep i\pi/2 \pm \alpha^\star$ with
\beq\label{alphastar}
	\cosh\alpha^\star =\sqrt{m_0/m},\quad \alpha>0
\eeq
and branch points on the same horizontal line in the $\theta$-plane, $\ep i\pi/2 \pm \alpha'$, but further away from the axis ${\rm Re}(\theta)=0$, at $\cosh\alpha' = m_0/m$, with $\alpha'>\alpha^\star$. Crossing symmetry \eqref{shift} implies $f^{\rho^{\sharp};\mu_+}_{+}(\alpha+i\pi/2) f^{\rho^{\sharp};\mu_-}_{-}(\alpha+i\pi/2) = - f^{\rho^{\sharp};\mu_+}_{-}(\alpha-i\pi/2) f^{\rho^{\sharp};\mu_-}_{+}(\alpha-i\pi/2)$. Then, using the fact that $U(\alpha+i\pi/2) = -U(\theta-i\pi/2)$ for $\alpha$ between $-\alpha'$ and $\alpha'$ (that is, away from the branch cut), we have
\[
	(1+U(\alpha+i\pi/2))f^{\rho^{\sharp};\mu_+}_{+}(\alpha+i\pi/2) f^{\rho^{\sharp};\mu_-}_{-}(\alpha+i\pi/2)
	=
	-
	(1-U(\alpha-i\pi/2))f^{\rho^{\sharp};\mu_+}_{-}(\alpha-i\pi/2) f^{\rho^{\sharp};\mu_-}_{+}(\alpha-i\pi/2).
\]
Hence, combining $\ep=\pm$, the integrals on $\alpha \in [-\alpha',\alpha']$ cancel out, and there only remain residue contributions at the poles $\alpha = \pm\alpha^\star$, and integrals along $[\alpha',\infty)$ and $[-\alpha',-\infty)$. The integrand in the latter integrals have exponential decaying factors $e^{-mx\cosh\alpha}$, which are subleading compared to the exponential factors $e^{-mx\cosh\alpha^\star}$ coming from the residues taken at the poles. The leading behavior is then determined by these residues,
\beqa
I_{\pm}&=&\pi i e^{-\sqrt{mm_0}x} f^{\rho^{\sharp};\mu_+}_{+}(\pi i/2+\theta_p) f^{\rho^{\sharp};\mu_-}_{-}(\pi i/2+\theta_p) \Res \left(U(z)\right)|_{z=\pi i/2\pm\alpha^\star}+\no\\
&&\pi i e^{-\sqrt{mm_0}x} f^{\rho^{\sharp};\mu_+}_{-}(-\pi i/2+\theta_p) f^{\rho^{\sharp};\mu_-}_{+}(-\pi i/2+\theta_p) \Res \left(U(z)\right)|_{z=-\pi i/2\pm\alpha^\star} \no
\eeqa
Using crossing symmetry again, we get:
\beq
I_{\pm}=e^{-\sqrt{mm_0}x} 4\pi \sqrt{m_0/m-1} f^{\rho^{\sharp};\mu_+}_{+}(\pi i/2+\theta_p) f^{\rho^{\sharp};\mu_-}_{-}(\pi i/2\pm\alpha^\star) \no
\eeq
Therefore, the leading behaviour is
\beq
	\t G(x) = O\lt(e^{-({\cal E}+\sqrt{mm_0})x}\rt)\quad
(m_0>m).
\eeq

On the other hand, if $m_0<m$, then $\sqrt{m_0/m}>m_0/m$, hence the branch points determine the least-decaying behavior. In this case, both poles and branch points are found on the imaginary axis below the ${\rm Im}(\theta) =\pi/2$ line. The branch points are at $\theta^\star = \pm i\kappa^\star$ with
\beq\label{kappastar}
	\sin\kappa^\star = m_0/m
\eeq
and the poles at $\pm i\kappa'$ with $\sin\kappa' = \sqrt{m_0/m}>\sin\kappa^\star$. Hence in this case, we shift the contours in the direction ${\rm sign}({\rm Im}(\theta)) = \ep$ up to the position of the poles at ${\rm Im}(\theta)=\ep i\kappa'$ (or a little bit before it), going around the part of the branch cut on ${\rm Im}(\theta)\in\ep [\kappa^*,\kappa']$. The integrals that remain on the lines $\theta = \ep i \kappa' + \alpha,\ \alpha\in\R$ have integrands with real-exponential factors $e^{-mx\sin\kappa' \cosh\alpha}$, hence are of order $e^{-mx\sin\kappa'}$. These are subleading compared to the integrals along ${\rm Im}(\theta)\in\ep [\kappa^*,\kappa']$, whose integrands have exponential factors $e^{-mx\sin\kappa}$ for $\kappa\in[\kappa^*,\kappa']$.

Taking only the integrals along the cut, with the change of variable $\theta = i\ep\kappa$, the leading behavior of the one-particle contribution is obtained from
\[
\sum_{\ep} \frc {i} 2 \int_{\kappa^\star}^{\kappa'} d\kappa \,e^{-mx\sin \kappa} (U(i\ep\kappa+{\bf 0}) - U(i\ep\kappa-{\bf 0})) f^{\rho^{\sharp};\mu_+}_{\ep}(i\ep\kappa) f^{\rho^{\sharp};\mu_-}_{-\ep}(i\ep\kappa).
\]
The leading large-$mx$ asymptotic behavior of this integral is obtained by linearizing around $\kappa=\kappa^\star$ the function in the exponential, taking the leading square-root form of the $U$-functions around that point, and replacing the form factors by their (finite) value at $\kappa=\kappa^\star$. The integral can then be extended to infinity. With $\kappa = \kappa^\star + \ell$, we obtain a quantity proportional to
\[
	e^{-mx\sin \kappa^\star} \int_{0}^\infty d\ell\,
	e^{-mx\ell \cos\kappa^\star} \sqrt{\ell} \propto (mx)^{-3/2} e^{-m_0 x}
\]
where we used \eqref{kappastar}. Hence in this case, the leading large-distance behavior is
\[
	\t G(x) = O\lt(x^{-3/2} e^{-({\cal E} + m_0)x}\rt)\quad
(m_0<m).
\]

In the ordered regime the leading behavior is simply the zero-particle contribution. The first subleading asymptotic terms are obtained by considering the two-particle form factor contributions,
\[
\sum_{\ep_1 \ep_2} \frac{\ep_1 \ep_2}{4} \int \frac{d\theta_1 d\theta_2}{2!} e^{\sum_{j=1}^{2}i\ep_j p_{\theta_j} x}U(\theta_1)U(\theta_2)f^{\rho^{\sharp};\sigma_+}_{\ep_1 \ep_2}(\theta_1,\theta_2) f^{\rho^{\sharp};\sigma_-}_{-\ep_2 -\ep_1}(\theta_2,\theta_1)
\]
Consider first $m_0>m$. Again, we deform the $\theta_1$ and $\theta_2$ contours in the same way as we did before. We first deform the $\theta_1$ contour up to ${\rm Im}(\theta_1)=\ep_1i\pi/2$. Again thanks to the crossing symmetry relation \eqref{shift} the integrals for $\ep_1=+$ and $\ep_1=-$ cancel out (for any $\ep_2$ and $\theta_2$), except for the residues of the poles at $\ep_1 i\pi/2 \pm \alpha^\star$ \eqref{alphastar} and for the branch cuts further away from the imaginary axis, on $|{\rm Re}(\theta)|>\alpha'$. The residues provide the part of the leading contribution, $O\lt(e^{-\sqrt{mm_0}x}\rt)$. We then deform the $\theta_2$ contour. The remaining integrals, after taking the $\theta_1$ residues at $\ep_1i\pi/2+s\alpha^\star$ (for $s=\pm$), are proportional to
\[
	\sum_{\ep_2} \ep_2 \int d\theta_2\,e^{i\ep_2 p_{\theta_2}x}
	U(\theta_2) h_{\ep_2}^{\sharp+}(\theta_2) h_{-\ep_2}^{\sharp-}(\theta_2)\tanh\lt(\frc{\theta_2 - (i\pi/2+s\alpha^\star)+i(\ep_2-1){\bf 0}}2\rt)^{2\ep_2}.
\]
Shifting to ${\rm Im}(\theta_2) = \ep_2 i\pi/2$, again the integrals for $\ep_2=+$ and $\ep_2=-$ cancel out, except for the poles at $\theta_2 = \ep_2 i\pi/2 \pm \alpha^\star$ and the branch cuts further away from the imaginary axis. For $\ep_2=+$, the pole of $U(\theta_2)$ at $\theta_2 = i\pi/2 +s \alpha^\star$ is cancelled by the zero in the $\tanh$ factor (similarly for the mirror pole with $\ep_2=-$). Hence only the pole at $\theta_2 = i\pi/2 -s \alpha^\star$ contributes. That is, overall, the leading large-distance behavior of the two-particle form factor contribution comes from the product of the residues taken at $\theta_1 = \ep_1 i\pi/2 + s\alpha^\star$ and $\theta_2 = \ep_2 i\pi/2 - s\alpha^\star$, with $s=+$ and $s=-$. By invariance under the exchange $\theta_1\leftrightarrow\theta_2$, the product of residues is independent of $s$. This then provides an overall decay $O\lt(e^{-2\sqrt{mm_0}s}\rt)$, so that we have
\[
	G(x) \propto e^{-{\cal E}x}\lt(1+O\lt(e^{-2\sqrt{mm_0}x}\rt)\rt)\quad
 (m_0>m).
\]

A similar analysis in the case $m_0>m$, where the branch cuts provide the leading behavior, gives
\[
	G(x) \propto e^{-{\cal E}x}\lt(1+O\lt(x^{-3}e^{-2m_0x}\rt)\rt)\quad
 (m_0<m).
\]

\section{Non-equilibrium steady-state form factor expansion}\label{appness}

We start with the expression (\ref{correlation}). In order to obtain convergent integrals, we shift towards imaginary directions. Let us consider one term in the sum $\sum_{\ep_1,\ldots,\ep_N}$. It is convenient to shift all $\theta_j$-contours with $\ep_j=+$ by a quantity $+i\pi$, and not to shift at all the integral contours with $\ep_k=-$. Using crossing symmetry (\ref{shift}) (more precisely, similar identities for all rapiditiez $\theta_j$), we see that on the shifted contours the integrand in (\ref{correlation}) becomes
\[
	(-1)^J
	\frac{e^{\sum_{j=1}^N \lt(ip_{\theta_j} x - iE_{\theta_j} t\rt)}}{\prod_{j=1}^{N}\lt(1-e^{-W(\theta_j)}\rt)}
f^{\rho^\sharp;\omega_{+}}_{-,\ldots,-}(\theta_1\cdots\theta_N) f^{\rho^\sharp; \omega_{-}}_{+,\ldots,+}(\theta_N\cdots \theta_1)
\]
where the number of minus signs $J$ is the number of shifted contours (the number of indices $j$ with $\ep_j=+$). Further, the shifting gives rise to residue contributions coming from the poles of the factors $(1-e^{-W_{\rm ness}(\theta_j)})^{-1}$, and to an integral running on both sides of the imaginary segment between 0 and $i\pi$ because the point $\theta_j=0$ is not analytic (discontinuity of $W_{\rm ness}(\theta_j)$). This means that we may replace every $\theta_j$-integral by
\[
	\int d\theta_j \mapsto \lt\{\ba{ll}
	\int d\theta_j & (\ep_j=-) \z
	-\int d\theta_j + \mbox{residues} + \mbox{imaginary segment}
	& (\ep_j=+).
	\ea\rt.
\]
Hence, under the sum over $\ep_1,\ldots,\ep_N$, only the residues and  imaginary segment contributions remain. By exchanging particles, we may assume that we take residues for the first $P$ particles and we integrate on imaginary segments for the remaining $Q=N-P$ particles, taking care of putting the extra combinatorial factor $N!/(P!Q!)$.

The pole in $\theta_j$ whose residue is taken is at position $i\pi/2+\alpha(n)$ for all $n\in\Z\setminus\{0\}$ (see (\ref{alphabeta})). Evaluating the residues and putting together the imaginary segment contributions, after a change to an integration over a real variable, one obtains (\ref{resness}).

In order to evaluate the leading large-$mx$ behavior (setting $t=0$ for simplicity) for the disorder two-point function, we consider $N=0$ and $M=1$, and only the part of the integral near $0$ and $\pi$ is sufficient. Hence we consider
\beq
	\frc{\bra\sigma\ket_{\rho_{\rm ness}}}2
	\,e^{-x{\cal E}_{\rm ness}} \int_0^\pi\, d\theta\,
	e^{-mx\sin\theta} \frc{\sinh \lt(\frc{\beta_l-\beta_r}2\cos\theta\rt)}{
	\sinh \lt(\frc{\beta_l}2\cos\theta\rt)
	\sinh \lt(\frc{\beta_r}2\cos\theta\rt)}
	{h^\sharp}_+^+(i\theta) \,{h^\sharp}_-^-(i\theta).
\eeq
The large-$mx$ behavior is evaluated by expanding for $\theta$ near to $0$ and $\pi$.  We can then use ${h^\sharp}_+^+(i\theta) \,{h^\sharp}_-^-(i\theta) \propto (i\theta)^{2i\gamma}$ near to $\theta=0$ from (\ref{smalltheta}), and similarly $\propto (i\pi - i\theta)^{-2i\gamma}$  near to $\theta=\pi$, using (\ref{pluspi}). Omitting the overall finite, real (temperature-dependent) factor, we then obtain, asymptotically,
\[
	e^{iB} \int_0^\infty d\theta\,e^{-mx\theta} \theta^{2i\gamma} + c.c.
	\propto \frc1{mx} \cos(2\gamma\log(mx)+B)
\]
for some phase $e^{iB}$. A more careful calculation gives (\ref{expmu}) with
\beqa
	A &=& 2\frc{\sinh\frc{(\beta_l-\beta_r)m}2}{\sinh\frc{\beta_l m}2
	\sinh\frc{\beta_rm}2} |\Gamma(1+2i\gamma)|\\
	B &=& {\rm arg}\big(\Gamma(1+2i\gamma)\big) +\n
	&& + \frc1{2\pi} \int_{|\theta|>1} d\theta\,
	\frc1{\sinh\theta}\log\coth\frc{W_{\rm ness}(\theta)}2
	+ \frc1{2\pi} \int_{|\theta|<1} d\theta\,
	\lt(\frc1{\sinh\theta}-\frc1\theta\rt)\log\coth\frc{W_{\rm ness}(\theta)}2.
	\no
\eeqa

Higher values of $M$ are likely to lead to contributions of exactly the same form. Indeed, we will have integrals of the type
\[
	\int_0^\pi d\theta_1d\theta_2\,\lt(\tan\frc{\theta_1-\theta_2}2\rt)^2
	e^{-mx(\sin\theta_1 + \sin\theta_2)}
	\times \mbox{leg factors}.
\]
For $\theta_1\sim0$ and $\theta_2\sim\pi$, and vice versa, the integrand has a second-order pole. This leads to logarithmic divergences, which we expect can be re-absorbed into the normalization of the field.


\begin{thebibliography}{99}
\bibitem{Karow} M. Karowski and P. Weisz: Exact form-factors in (1+1)-dimensional field theoretic models with soliton behaviour, Nucl. Phys. B139, 455 (1978).
\bibitem{Zamo} A. B. Zamolodchikov and Al. B. Zamolodchikov: Factorized S-matrices in two dimensions as the exact solutions of certain relativistic quantum field theory models, Ann. Phys. 120, 253 (1979).
\bibitem{Smirnov} F. Smirnov: Form factors in completely integrable models of quantum field theory, Adv. Series in Math. Phys. 14, World Scientific, Singapore (1992).
\bibitem{Delfino} G. Delfino: Integrable field theory and critical phenomena: the Ising model in a magnetic field, J. Phys. A 37, R45 (2004), \href{http://arxiv.org/abs/hep-th/0312119}{arXiv:hep-th/0312119v1}\;.
 \bibitem{Mussardo} G. Mussardo: Statistical Field Theory: An Introduction to Exactly Solved Models in Statistical Physics, Oxford University Press, New York (2010).
  \bibitem{Ess} F. H. L. Essler and R. M. Konik: Applications of massive integrable quantum field theories to problems in condensed matter physics, in: From Fields to Strings: Circumnavigating Theoretical Physics, Ian Kogan Memorial Collection, Volume 1, ed. Misha Shifman, Arkady Vainshtein and John Wheater, World Scientific (2004).
\bibitem{ZamoLee} Al.B. Zamolodchikov: Two-point correlation function in scaling Lee-Yang model, Nucl. Phys. B 348, 619 (1991).
\bibitem{Yurov} V. P. Yurov and Al. B. Zamolodchikov: Correlation functions of integrable 2D models of relativistic field theory: Ising model, Int. J. Mod. Phys. 6, 3419 (1991). 
 \bibitem{Rigol} M. Rigol, V. Dunjko, V. Yurovsky and M. Olshanii: Relaxation in a Completely Integrable Many-Body Quantum
System: An Ab Initio Study of the Dynamics of the Highly Excited States of Lattice Hard-Core Bosons, Phys.
Rev. Lett. 98, 50405 (2007), \href{http://arxiv.org/pdf/cond-mat/0604476v2.pdf}{arXiv:cond-mat/0604476v2}\;.
\bibitem{Rigol2} M. Rigol, V. Dunjko, and M. Olshanii: Thermalization and its mechanism for generic isolated quantum systems, Nature 452, 854 (2008), \href{http://arxiv.org/pdf/0708.1324v2.pdf}{arXiv:0708.1324v2}\;.
 \bibitem{Pol} A. Polkovnikov, K. Sengupta, A. Silva and M. Vengalattore: Nonequilibrium dynamics of closed interacting quantum systems, Rev. Mod. Phys. 83, 863 (2011), \href{http://arxiv.org/abs/1007.5331}{arXiv:1007.5331v2 [cond-mat.stat-mech]}\;.
  \bibitem{Fagotti} P. Calabrese, F.H.L. Essler and M. Fagotti: Quantum Quench in the Transverse Field Ising Chain, Phys. Rev. Lett. 106, 227203 (2011), \href{http://arxiv.org/pdf/1104.0154v1.pdf}{arXiv:1104.0154v1 }.
    \bibitem{Fagotti1} P. Calabrese, F. H. L. Essler and M. Fagotti: Quantum Quench in the Transverse Field Ising chain I: Time evolution
of order parameter correlators, J. Stat. Mech. (2012) P07016, \href{http://arxiv.org/pdf/1204.3911v3.pdf}{arXiv:1204.3911v3}\;.
    \bibitem{Aschbacher2} W. H. Aschbacher and C.-A. Pillet: Non-equilibrium steady states of the XY Chain, J.
Stat. Phys. 112, 1153  (2003).
   \bibitem{Ben4}  D. Bernard and B. Doyon:  Energy flow in non-equilibrium conformal field theory, J. Phys. A: Math. Theor. 45, 362001 (2012), \href{http://arxiv.org/pdf/1202.0239.pdf}{arXiv:1202.0239}.
   \bibitem{BDaihp} D. Bernard and B. Doyon, Non-equilibrium steady states in conformal field theory, to appear in Ann. Henri Poincar\'e (2014), \href{http://arxiv.org/pdf/1302.3125v2.pdf}{arXiv:1302.3125}\;.
   \bibitem{Ben2} B. Doyon: Nonequilibrium density matrix for thermal transport in quantum field theory, lecture notes, Les Houches School on Strongly Interacting Quantum Systems Out of Equilibrium (2012), \href{http://arxiv.org/pdf/1212.1077.pdf}{arXiv:1212.1077v1}.
   
\bibitem{DeLuca} A. De Luca, J. Viti, D. Bernard and B. Doyon: Non-equilibrium thermal transport in the quantum Ising chain, Phys. Rev. B 88, 134301 (2013), \href{http://arxiv.org/pdf/1305.4984v2.pdf}{arXiv:1305.4984}\;.
  
   \bibitem{Kapusta} J. I. Kapusta: Finite Temperature Field Theory, Cambridge University Press, Cambridge, 1989.
   
   \bibitem{Leclair} A. Leclair, F. Lesage, S. Sachdev and H. Saleur: Finite temperature correlations in the one-dimensional quantum Ising model, Nucl.Phys. B482 (1996) 579-612 ,\href{http://arxiv.org/pdf/cond-mat/9606104v2.pdf}{arXiv:cond-mat/9606104v2}\;.






\bibitem{Sachdev1} S. Sachdev and A. P. Young: Low temperature relaxational dynamics of the Ising chain in a transverse field
Subir Sachdev, Phys. Rev. Lett. 78, 2220 (1997), \href{http://arxiv.org/pdf/cond-mat/9609185v2.pdf}{arXiv:cond-mat/9609185v2}\;.

\bibitem{Sachdev2} K. Damle and S. Sachdev: Non-zero temperature transport near quantum critical points, Rev. B 56, 8714 (1997), \href{http://arxiv.org/pdf/cond-mat/9705206v3.pdf}{arXiv:cond-mat/9705206v3}\;.

\bibitem{Leclair2} A. Leclair and G. Mussardo: Finite temperature correlation functions in integrable QFT, Nucl. Phys. B 552, 624 (1999), \href{http://arxiv.org/pdf/hep-th/9902075v1.pdf}{arXiv:hep-th/9902075v1}\;.

\bibitem{Saleur} H. Saleur: A comment on finite temperature correlations in integrable QFT, Nucl. Phys. B 567, 602 (2000), \href{http://arxiv.org/pdf/hep-th/9909019v1.pdf}{arXiv:hep-th/9909019v1}\;.

\bibitem{Mussardo2} G. Mussardo: On the finite temperature formalism in integrable quantum field theories, J. Phys. A 34, 7399 (2001), \href{http://arxiv.org/pdf/hep-th/0103214v2.pdf}{arXiv:hep-th/0103214}\;.

\bibitem{Konik} R. Konik: Haldane gapped spin chains: exact low temperature expansions of correlation functions, Phys. Rev. B 68, 104435 (2003), \href{http://arxiv.org/pdf/cond-mat/0105284v1.pdf}{arXiv:cond-mat/0105284}\;.

\bibitem{Fonseca1} P. Fonseca and A. B. Zamolodchikov: Ward identities and integrable differential equations in the Ising field theory, \href{http://arxiv.org/pdf/hep-th/0309228v1.pdf}{hep-th/0309228}\;.

   \bibitem{Fonseca2} P. Fonseca and A. B. Zamolodchikov: Ising field theory in a magnetic field: analytic properties of the free energy, J. Stat. Phys. 110, 527 (2003), \href{http://arxiv.org/pdf/hep-th/0112167v1.pdf}{arXiv:hep-th/0112167}\;.

  \bibitem{Ben1} B. Doyon: Finite-temperature form factors in the free Majorana theory, J. Stat. Mech. (2005) P11006, \href{http://arxiv.org/pdf/hep-th/0506105.pdf}{hep-th/0506105}\;.

\bibitem{Altshuler} B. L. Altshuler, R. M. Konik and A. M. Tsvelik: Low temperature correlation functions in integrable models: derivation of the large distance and time asymptotics from the form factor expansion, Nucl. Phys. B 739, 311 (2006), \href{http://arxiv.org/pdf/cond-mat/0508618v3.pdf}{arXiv:cond-mat/0508618v3}\;.












\bibitem{Reyes06} S. A. Reyes and A. M. Tsvelik: Finite-temperature correlation function for the one-dimensional quantum Ising model: the virial expansion, Phys. Rev. B 73, 220405(R) (2006), \href{arXiv:cond-mat/0605040v1}{arXiv:cond-mat/0605040v1}\;.

\bibitem{Damle05} K. Damle and S. Sachdev: Universal relaxational dynamics of gapped one-dimensional models in the quantum sine-Gordon universality class, Phys. Rev. Lett. 95, 187201 (2005), \href{http://arxiv.org/pdf/cond-mat/0507380v2.pdf}{arXiv:cond-mat/0507380v2}\;.

\bibitem{Rapp} A. Rapp and G. Zarand: Dynamical correlations and quantum phase transition in the quantum Potts model, Phys. Rev. B 74, 014433 (2006), \href{http://arxiv.org/pdf/cond-mat/0507390v6.pdf}{arXiv:cond-mat/0507390v6}\;.

   \bibitem{Ben3} B. Doyon: Finite-temperature form factors: a review, SIGMA 3, 011 (2007), \href{http://arxiv.org/pdf/hep-th/0611066.pdf}{hep-th/0611066}, in: Proceedings of the O'Raifeartaigh Symposium on Non-Perturbative and Symmetry Methods in Field Theory� (Budapest, 2006).

\bibitem{Gamsa} B. Doyon and A. Gamsa: Integral equations and long-time asymptotics for finite-temperature Ising chain correlation functions, J. Stat. Mech. (2008) P03012, \href{http://arxiv.org/pdf/0711.4619v1.pdf}{arXiv:0711.4619v1}\;.

\bibitem{Pozsgay1} B. Pozsgay and G. Tak\'acs: Form factors in finite volume I: form factor bootstrap and truncated conformal space, Nucl. Phys. B 788, 167 (2008), \href{http://arxiv.org/pdf/0706.1445v2.pdf}{arXiv:0706.1445v2}\;.

\bibitem{Pozsgay2} B. Pozsgay and G. Tak\'acs: Form factors in finite volume II: disconnected terms and finite temperature correlators, Nucl. Phys. B 788, 209 (2008), \href{http://arxiv.org/pdf/0706.3605v1.pdf}{arXiv:0706.3605v1}\;.

\bibitem{EssKon1} F. H. L. Essler and R. M. Konik: Finite-temperature lineshapes in gapped quantum spin chains, Phys. Rev.
B78,  100403 (2008), \href{http://arxiv.org/pdf/0711.2524v2.pdf}{arXiv:0711.2524v2}\;.

\bibitem{EssKon2} F. H. L. Essler and R. M. Konik: Finite-temperature dynamical correlations in massive integrable quantum field theories, J. Stat. Mech. 0909
(2009) P09018, \href{http://arxiv.org/pdf/0907.0779v2.pdf}{arXiv:0907.0779v2}\;.

\bibitem{Pozsgay3} B. Pozsgay and G. Takacs: Form factor expansion for thermal correlators, J. Stat. Mech. (2010) P11012, \href{http://arxiv.org/pdf/1008.3810v3.pdf}{arXiv:1008.3810v3}\;.


\bibitem{Sz} I.M. Sz\'ecs\'enyi and G. Tak\'acs: Spectral expansion for finite temperature two-point functions and clustering, \href{http://arxiv.org/pdf/1210.0331v1.pdf}{arXiv:1210.0331v1}\;.


   
   \bibitem{Fagotti2} P. Calabrese, F. H. L. Essler and M. Fagotti:  Quantum Quench in the Transverse Field Ising Chain II: Stationary State Properties, J. Stat. Mech. (2012) P07022, \href{http://arxiv.org/pdf/1205.2211v2.pdf}{arXiv:1205.2211v2}\;.
   
   \bibitem{MusQuench} G. Mussardo: Infinite-time average of local fields in an integrable quantum field theory after a quantum quench, Phys. Rev. Lett. 111, 100401 (2013), \href{http://arxiv.org/pdf/1308.4551v1.pdf}{arXiv:1308.4551}\;.
   \bibitem{Aschbacher} W. H. Aschbacher and J.-M. Barbaroux: Out of equilibrium correlations in the XY chain, Lett. Math. Phys. 77, 11 (2006), \href{arXiv:math-ph/0505062v2}{arXiv:math-ph/0505062v2}\;.
   
   \bibitem{Fano1} U. Fano: Description of states in quantum mechanics by density matrix and operator techniques, Rev. Mod. Phys. 29, 74 (1957).
   \bibitem{Crawford} J. A. Crawford: An alternative method of quantization: the existence of classical fields, Nuovo Cim. 10, 698 (1958).
   \bibitem{Fano2} U. Fano: Liouville Representation of Quantum Mechanics with Application to Relaxation Processes, in Lectures on the Many-Body Problem, ed. E. R. Caianiello, Academic Press, New York, p. 217 (1964)
   \bibitem{Schmutz} M. Schmutz: Real-time Green's functions in many-body problems, Z. Physik B 30, 97 (1978).
   \bibitem{Umezawa} H. Umezawa, H. Matsumoto and M. Tachiki: Thermo Field Dynamics and Condensed States, North-Holland, Amsterdam, 1982.

\bibitem{Arimitsu} T. Arimitsu and H. Umezawa: Non-equilibrium Thermo Field Dynamics, Prog. Theo. Phys. 77, 32 (1987)

\bibitem{Gelfand} I. M. Gelfand and M. A. Naimark: On the imbedding of normed rings into the ring of operators on a Hilbert space, Math. Sbornik 12, 197  (1943).

\bibitem{Segal}  I. E. Segal: Irreducible representations of operator algebras, Bull. Am. Math. Soc. 53, 73 (1947).

  \bibitem{Arveson} W. Arveson: An Invitation to C*-Algebra, Springer-Verlag, 1981.


\bibitem{Bugrij} A. I. Bugrij: Form Factor Representation of the Correlation Functions of the two Dimensional Ising Model on a Cylinder, in Integrable Structures of Exactly Solvable Two-Dimensional Models of Quantum Field Theory, ed. S. Pakuliak. and G. Gehlen, NATO Science Series, Springer Netherlands, Volume 35, p. 65, 2001.

\bibitem{Bugrij2} A. I. Bugrij: Correlation function of the two-dimensional Ising model on the finite lattice. I, Theor. Math. Phys. 127, 528 ,(2001).

\bibitem{Bugrij3} A. I. Bugrij and O. Lisovyy: Magnetic susceptibility of the two-dimensional Ising model on a finite lattice, JETP 94, 1140 (2002).

\bibitem{Bugrij4} A. I. Bugrij and O. Lisovyy: Spin matrix elements in 2D Ising model on the finite lattice, Phys. Lett. A 319, 390 (2003), \href{http://arxiv.org/pdf/0708.3625v1.pdf}{arXiv:0708.3625v1}\;.

\bibitem{Bugrij5} A. I. Bugrij and O. Lisovyy: Correlation function of the two-dimensional Ising model on a finite lattice. II, Theor. Math. Phys. 140, 987 (2004), \href{http://arxiv.org/pdf/0708.3643v1.pdf}{arXiv:0708.3643v1}\;.

\bibitem{Iorgov} N. Iorgov and O. Lisovyy: Finite-lattice form factors in free-fermion models, J. Stat. Mech. (2011) P04011, \href{http://arxiv.org/pdf/1102.2145v2.pdf}{arXiv:1102.2145v2}\;.

 \bibitem{Iorgov2} N. Iorgov: Form-factors of the finite quantum XY-chain, J. Phys. A 44, 335005 (2011),  \href{http://arxiv.org/pdf/0912.4466.pdf}{arXiv:0912.4466v2}\;.

\bibitem{Schuricht2012} D. Schuricht and F. H. L. Essler: Dynamics in the Ising field theory after a quantum quench, J. Stat. Mech. (2012) P04017, \href{http://arxiv.org/pdf/1203.5080v2.pdf}{arXiv:1203.5080}\;.



\bibitem{Lieb61} E. Lieb, T. Schultz and D. Mattis: Two soluble models of an antiferromagnetic chain, Ann. Physics 16, 407 (1961).
\bibitem{Katsura62} S. Katsura: Statistical Mechanics of the Anisotropic Linear Heisenberg Model, Phys. Rev. 127, 1508 (1962).
\bibitem{JordanWigner} P. Jordan and W. Wigner: \"Uber das Paulische \"Aquivalenzverbot, Z. Physik 47, 631 (1928).
\bibitem{McCoy68} B. McCoy: Spin Correlation Functions of the X-Y Model, Phys. Rev. 173, 531 (1968).
\bibitem{Pfeuty70} P. Pfeuty: The one-dimensional Ising model with a transverse field, Ann. Phys. (N. Y.) 57, 79 (1970).
\bibitem{BarouchMcCoyII} E. Barouch and B.M. McCoy:
Statistical mechanics of the XY model. II. Spin-correlation functions, Phys.Rev. A 3, 786 (1971).
  \bibitem{Belavin} A. A. Belavin, A. M. Polyakov and A. B. Zamolodchikov: Infinite conformal symmetry in two dimensional
quantum field theory, Nucl. Phys. B241, 333 (1984).
 \bibitem{Itzykson} C. Itzykson and J.�M. Drouffe:
Statistical field theory, Cambridge University Press, 1989
\bibitem{Berg} B. Berg, M. Karowski and P. Weisz: Construction of Green's functions from an exact S-matrix. Phys. Rev. D19, 2477 (1979).
   \bibitem{Kin} T. Kinoshita, T. Wenger and D. S. Weiss: A quantum Newton's cradle, Nature 440, 900 (2006).
   
   \bibitem{CAetal} O. Castro-Alvaredo, Y. Chen, B. Doyon and M. Hoogeveen: Thermodynamic Bethe ansatz for non-equilibrium steady states: exact energy current and fluctuations in integrable QFT, to appear in J. Stat. Mech. (2014), \href{http://arxiv.org/pdf/1310.4779v1.pdf}{arXiv:1310.4779}\;.
   \bibitem{Babelon} O. Babelon, D. Bernard and M. Talon: Introduction to classical integrable systems, Cambridge University Press, 2003

\bibitem{Babelon2}  O. Babelon and D. Bernard: From Form Factors to Correlation Functions: The Ising Model, Phys. Lett. B 288, 113 (1992), \href{http://arxiv.org/pdf/hep-th/9206003v1.pdf}{arXiv:hep-th/9206003v1}\;.


  \bibitem{Izergin} A. G. Izergin, V. S. Kapitonov and N. A.  Kitanine: Equal-time temperature correlators of the one-dimensional Heisenberg XY chain, Zap. Nauchn. Semin. POMI 245, 173  (1997), \href{http://arxiv.org/pdf/solv-int/9710028v1.pdf}{arXiv:solv-int/9710028v1}\;.






\end{thebibliography}
\end{document}